\documentclass[aps,prd,preprint,groupedaddress]{revtex4}
\usepackage{array,hhline,dcolumn} 
\usepackage{rotating} 

\newcommand{\gtwid}{\mathrel{\raise.3ex\hbox{$>$\kern-.75em\lower1ex\hbox{$\sim$}}}}
\newcommand{\ltwid}{\mathrel{\raise.3ex\hbox{$<$\kern-.75em\lower1ex\hbox{$\sim$}}}}

\begin{document}
%
\title{Updated MiniBooNE neutrino oscillation results with increased data and new background studies}

\author{
        A.~A. Aguilar-Arevalo$^{14}$,
        B.~C.~Brown$^{5}$, 
        J.~M.~Conrad$^{13}$,
        R.~Dharmapalan$^{1,7}$, 
        A.~Diaz$^{13}$,
        Z.~Djurcic$^{2}$, 
        D.~A.~Finley$^{5}$, 
        R.~Ford$^{5}$,
        G.~T.~Garvey$^{10}$,
        S.~Gollapinni$^{10}$,
        A.~Hourlier$^{13}$, 
        E.-C.~Huang$^{10}$,
        N.~W.~Kamp$^{13}$, 
        G.~Karagiorgi$^{4}$, 
        T.~Katori$^{12}$,
        T.~Kobilarcik$^{5}$, 
        K.~Lin$^{4,10}$,
        W.~C.~Louis$^{10}$, 
        C.~Mariani$^{17}$, 
        W.~Marsh$^{5}$,
        G.~B.~Mills$^{10,\dagger}$,
        J.~Mirabal-Martinez$^{10}$, 
        C.~D.~Moore$^{5}$, 
        R.~H.~Nelson$^{3,\star}$, 
        J.~Nowak$^{9}$,
        I.~Parmaksiz$^{16}$, 
        Z.~Pavlovic$^{5}$, 
        H.~Ray$^{6}$, 
        B.~P.~Roe$^{15}$,
        A.~D.~Russell$^{5}$,
	A.~Schneider$^{13}$,
        M.~H.~Shaevitz$^{4}$,
	H.~Siegel$^{4}$,
        J.~Spitz$^{15}$, 
        I.~Stancu$^{1}$,
	R.~Tayloe$^{8}$,
        R.~T.~Thornton$^{10}$, 
        M.~Tzanov$^{3,11}$,
        R.~G.~Van~de~Water$^{10}$,
        D.~H.~White$^{10,\dagger}$, 
        E.~D.~Zimmerman$^{3}$ \\
\smallskip
(The MiniBooNE Collaboration)
\smallskip
}
\smallskip
\smallskip
\affiliation{
$^1$University of Alabama; Tuscaloosa, AL 35487, USA \\
$^2$Argonne National Laboratory; Argonne, IL 60439, USA \\
$^3$University of Colorado; Boulder, CO 80309, USA \\
$^4$Columbia University; New York, NY 10027, USA \\
$^5$Fermi National Accelerator Laboratory; Batavia, IL 60510, USA \\
$^6$University of Florida; Gainesville, FL 32611, USA \\
$^7$University of Hawaii, Manoa; Honolulu, HI 96822, USA \\
$^8$Indiana University; Bloomington, IN 47405, USA \\
$^9$Lancaster University; Lancaster LA1 4YB, UK \\
$^{10}$Los Alamos National Laboratory; Los Alamos, NM 87545, USA \\
$^{11}$Louisiana State University; Baton Rouge, LA 70803, USA \\
$^{12}$King's College London; London WC2R 2LS, UK \\
$^{13}$Massachusetts Institute of Technology; Cambridge, MA 02139, USA \\
$^{14}$Instituto de Ciencias Nucleares; Universidad Nacional Aut\'onoma de M\'exico; CDMX 04510, M\'exico \\
$^{15}$University of Michigan; Ann Arbor, MI 48109, USA \\
$^{16}$University of Texas at Arlington, Arlington, TX 76019 \\
$^{17}$Center for Neutrino Physics; Virginia Tech; Blacksburg, VA 24061, USA \\
$^\star$Now at The Aerospace Corporation, Los Angeles, CA 90009, USA \\
$^\dagger$Deceased \\
}

\date{\today}

\begin{abstract}
The MiniBooNE experiment at Fermilab reports a total excess of $638.0 \pm 52.1(stat.) \pm 122.2(syst.)$ electronlike
events from a data sample corresponding to $18.75 \times 10^{20}$
protons-on-target in neutrino mode, which is a 46\% increase in the data sample
with respect to previously published results, and 
$11.27 \times 10^{20}$ protons-on-target in antineutrino mode. 
The overall significance of the excess, $4.8 \sigma$, is limited by systematic uncertainties, 
assumed to be Gaussian, as the statistical significance of
the excess is $12.2 \sigma$.
The additional statistics allow several studies to address questions on the source of the excess.   
First, we provide two-dimensional plots in visible energy and cosine of the angle of the outgoing lepton, 
which can provide valuable input to models for the event excess.   
Second, we test whether the excess may arise from photons that enter the detector from external events
or photons exiting the detector from $\pi^0$ decays in two model independent ways. 
Beam timing information shows that almost all of the excess is in time with neutrinos that interact 
in the detector. The radius distribution shows that the excess is distributed throughout the volume, 
while tighter cuts on the fiducial volume increase the significance of the excess. 
The data likelihood ratio disfavors models that
explain the event excess due to entering or exiting photons.
\end{abstract}

\pacs{14.60.Lm, 14.60.Pq, 14.60.St}

\keywords{Suggested keywords}
\maketitle


\section{Introduction}

The LSND \cite{lsnd} and MiniBooNE \cite{mb_oscnew,mb_oscnst} experiments have reported excesses of
$\nu_e$ and $\bar \nu_e$ charge-current quasielastic (CCQE) events in $\nu_\mu$ beams. 
Exotic models beyond the three-neutrino paradigm that have been invoked to explain these anomalies include, 
for example, 3+N neutrino oscillation models 
involving three active neutrinos and N additional sterile neutrinos 
\cite{sorel,karagiorgi,collin,giunti,giunti2,giunti3,kopp,kopp2,white_paper,3+2,diaz}, 
resonant neutrino oscillations \cite{resonant}, Lorentz violation 
\cite{lorentz}, sterile neutrino decay \cite{sterile_decay}, scalar decay \cite{scalar_decay},
sterile neutrino nonstandard interactions
\cite{NSI}, and altered dispersion relations with sterile neutrinos \cite{sterile_extra}.
This paper presents improved MiniBooNE $\nu_e$ appearance results with increased statistics and with
additional studies that disfavor neutral-current (NC) $\pi^0$ and external event backgrounds.

\section{The MiniBooNE Experiment}

\begin{figure}[tbp]
\vspace{+0.1in}
\centerline{\includegraphics[angle=0, width=9.0cm]{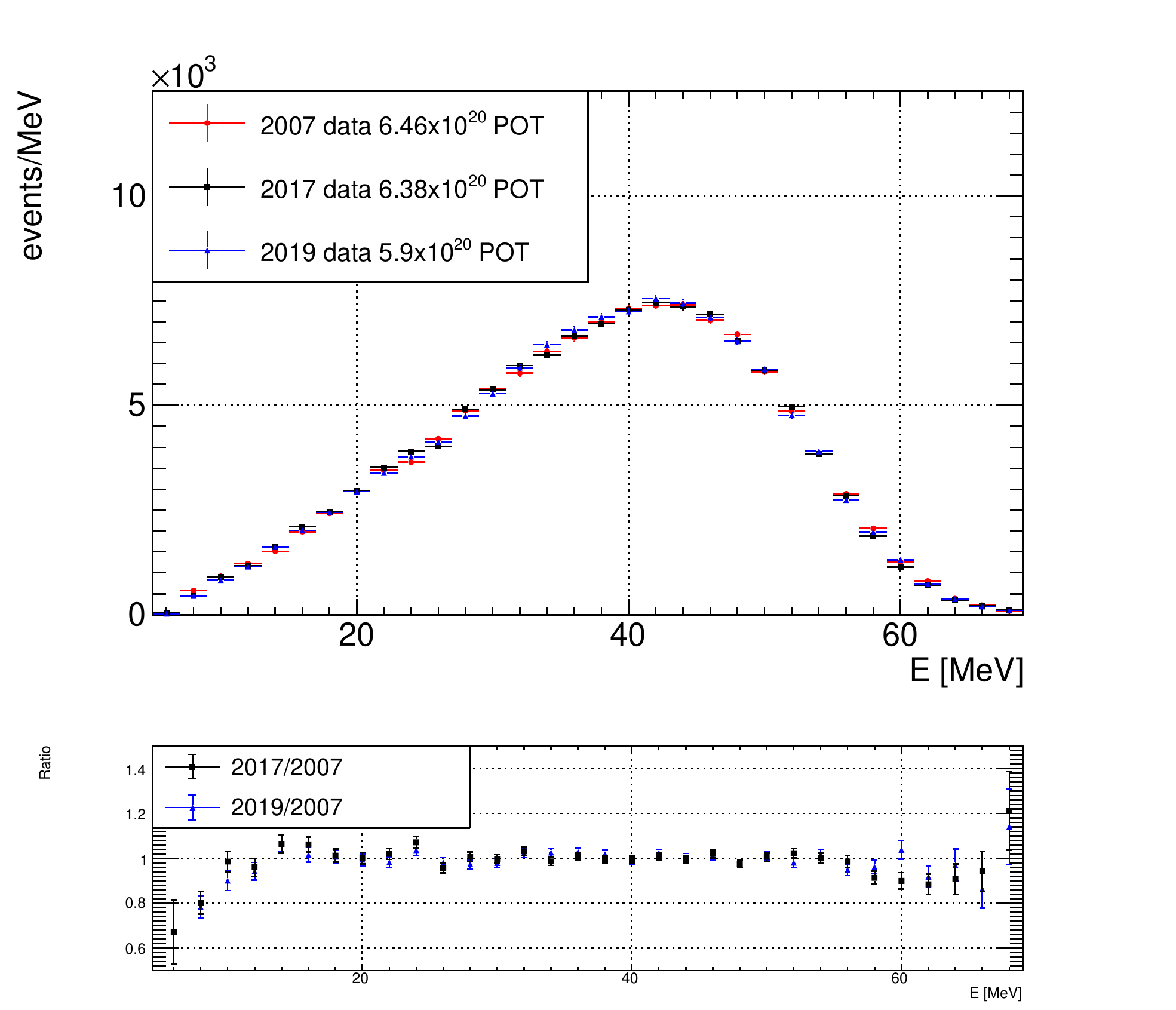}}
\vspace{-0.2in}
\caption{The Michel electron energy distribution
for the first, second, and third running periods in neutrino mode.
The events are normalized to the first running period. The bottom plot shows ratios of
the second and third running periods to the first running period.}
\label{Michel}
\vspace{0.1in}
\end{figure}

The MiniBooNE experiment makes use of the Booster Neutrino Beam (BNB) that
is produced by 8 GeV protons from
the Fermilab Booster interacting on a beryllium target inside a magnetic focusing horn,
followed by meson decay in a 50 m decay pipe.
In neutrino mode, the $\nu_\mu$, $\bar \nu_\mu$, $\nu_e$, and $\bar \nu_e$ flux
contributions at the detector are 93.5\%, 5.9\%, 0.5\%, and 0.1\%, respectively, while
in antineutrino mode, the flux
contributions are 15.7\%, 83.7\%, 0.2\%, and 0.4\%, respectively.
The $\nu_\mu$ and $\bar{\nu}_{\mu}$ fluxes peak at approximately 600 MeV and 400 MeV, respectively. 
The MiniBooNE detector, described in detail in reference \cite{mb_detector}, consists of
a 12.2 m diameter sphere filled with 818 tonnes of pure mineral oil (CH$_{2}$) and
is located 541 m from the beryllium target. The detector is covered by 1520 8-inch
photomultiplier tubes (PMTs), where 1280 PMTs are in the interior detector region and 240
PMTs are located in the optically isolated outer veto region. The PMTs detect the directed
Cherenkov light and the isotropic scintillation light produced by charged particles from neutrino
interactions in the mineral oil. Events are reconstructed \cite{mb_recon} from
the hit PMT charge and time information, and the reconstructed neutrino energy, $E_\nu^{QE}$, 
is estimated from the measured energy and angle of the outgoing muon 
or electron, assuming the kinematics of CCQE scattering \cite{ccqe}.
The MiniBooNE experiment has collected data from 2002-2019, based on a total of $11.27 \times 10^{20}$ 
protons-on-target (POT) in antineutrino mode and $18.75 \times 10^{20}$ POT
in neutrino mode. Also, a special
beam off-target run collected an additional $1.86 \times 10^{20}$ POT
in a search for sub-GeV dark matter \cite{mb_dm}. 
During the 17 years of
running, the BNB and MiniBooNE detector have been stable to within
3\% in neutrino energy. Fig. \ref{Michel} shows the energy distribution of Michel electrons from
stopped muon decay for the first ($6.46 \times 10^{20}$ POT from 2002 to 2007), second ($6.38 \times 10^{20}$ POT
from 2015 to 2017), 
and third running periods ($5.91 \times 10^{20}$ POT from 2017 to 2019) in neutrino mode.
By adjusting the energy calibration by 2\% for the second running period and by 3\% for the third
running period, good agreement is obtained for the Michel electron energy distribution.

\section{Data Analysis}

\begin{figure}[tbp]
\vspace{+0.1in}
\centerline{\includegraphics[angle=0, width=9.0cm]{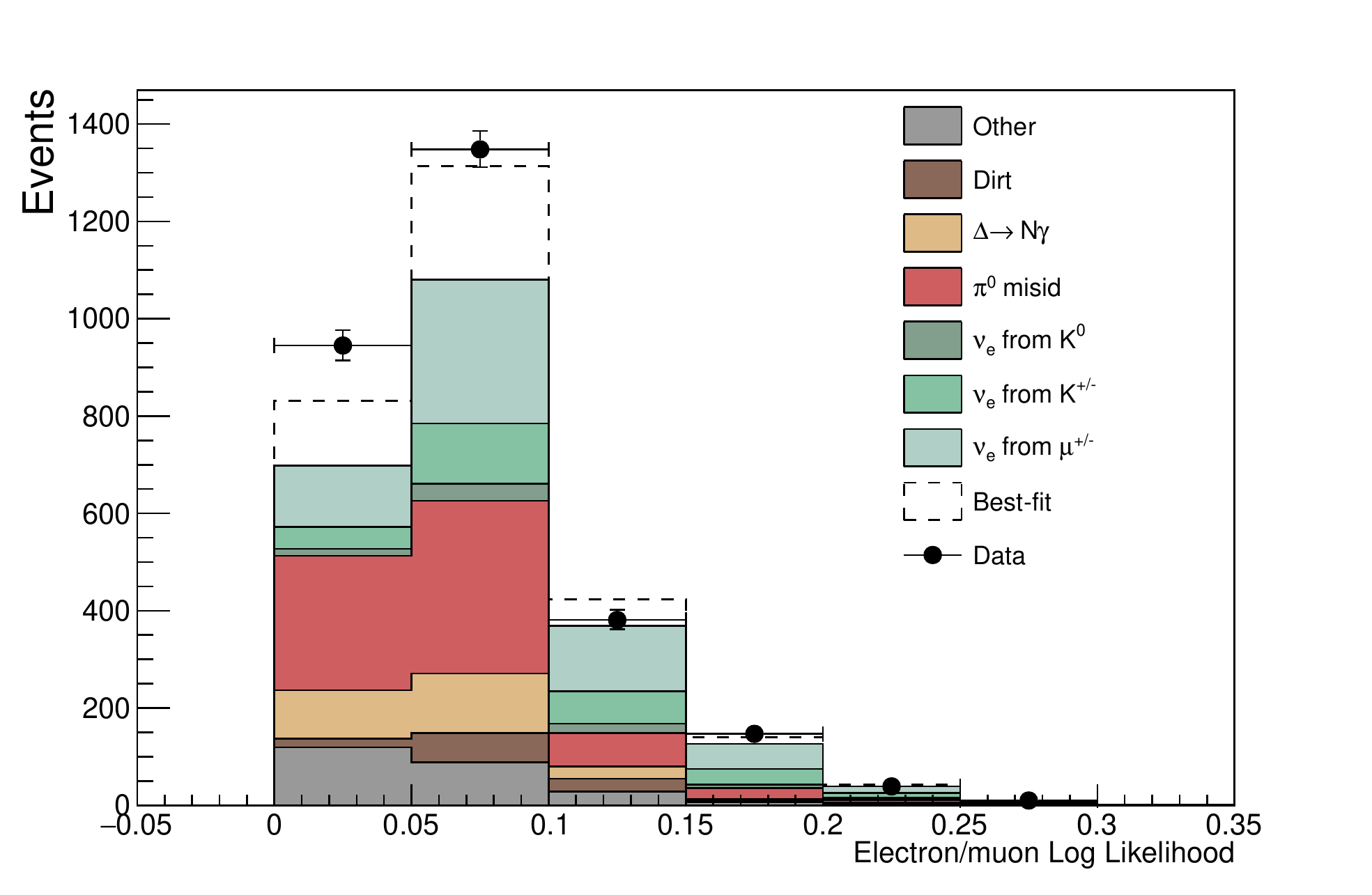}}
\vspace{-0.2in}
\caption{The MiniBooNE neutrino mode
electron-muon particle identification distributions, corresponding to the total $18.75 \times 10^{20}$ POT data
in the $200<E_\nu^{QE}<1250$ MeV energy range,
for ${\nu}_e$ CCQE data (points with statistical errors) and background (colored histogram).
The dashed histogram shows the best fit to the neutrino-mode data assuming two-neutrino oscillations.}
\label{PIDem}
\vspace{0.1in}
\end{figure}

\begin{figure}[tbp]
\vspace{+0.1in}
\centerline{\includegraphics[angle=0, width=9.0cm]{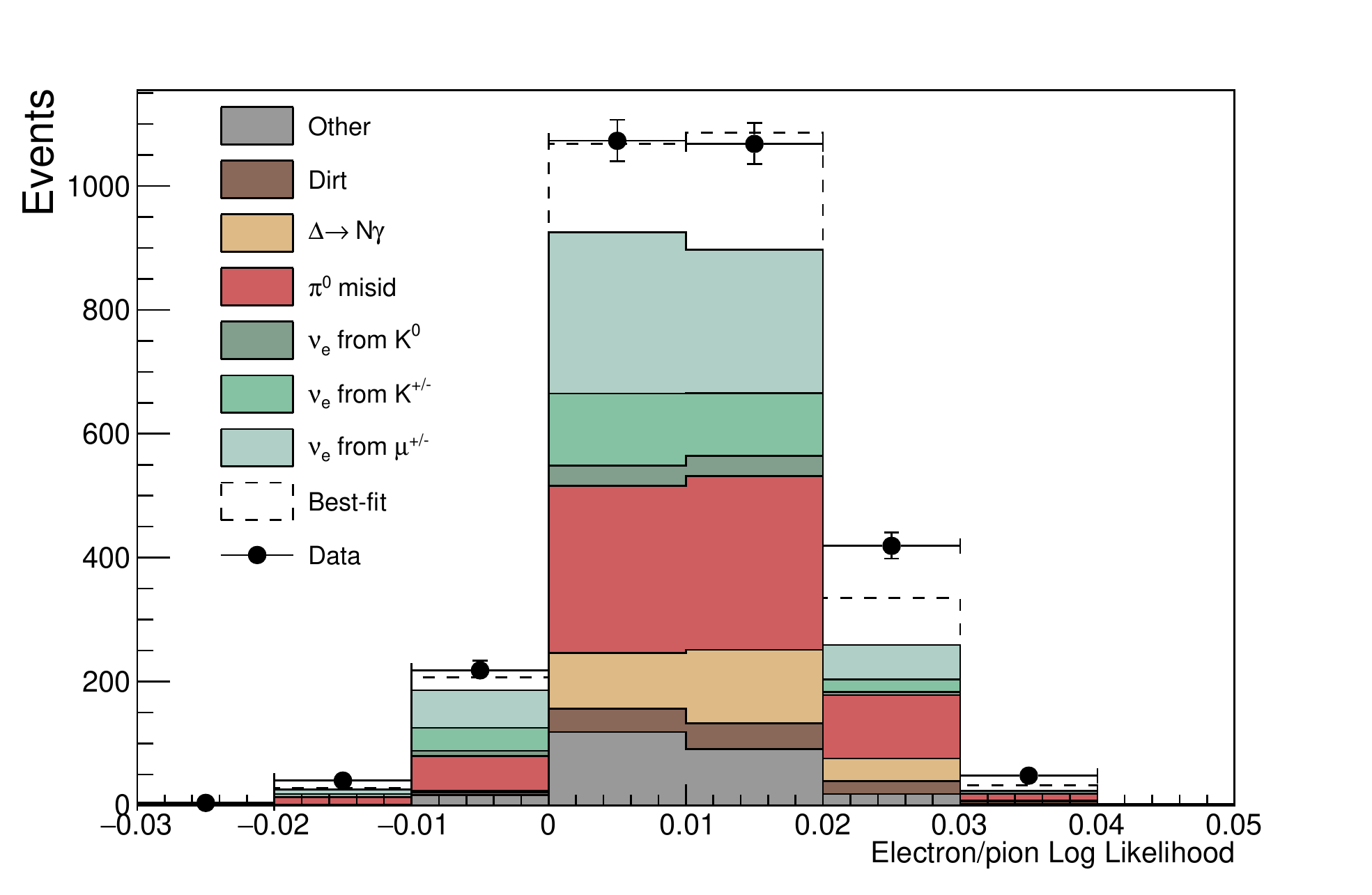}}
\vspace{-0.2in}
\caption{The MiniBooNE neutrino mode
electron-pion particle identification distributions, corresponding to the total $18.75 \times 10^{20}$ POT data
in the $200<E_\nu^{QE}<1250$ MeV energy range,
for ${\nu}_e$ CCQE data (points with statistical errors) and background (colored histogram).
The dashed histogram shows the best fit to the neutrino-mode data assuming two-neutrino oscillations.}
\label{PIDep}
\vspace{0.1in}
\end{figure}

\begin{figure}[tbp]
\vspace{+0.1in}
\centerline{\includegraphics[angle=0, width=9.0cm]{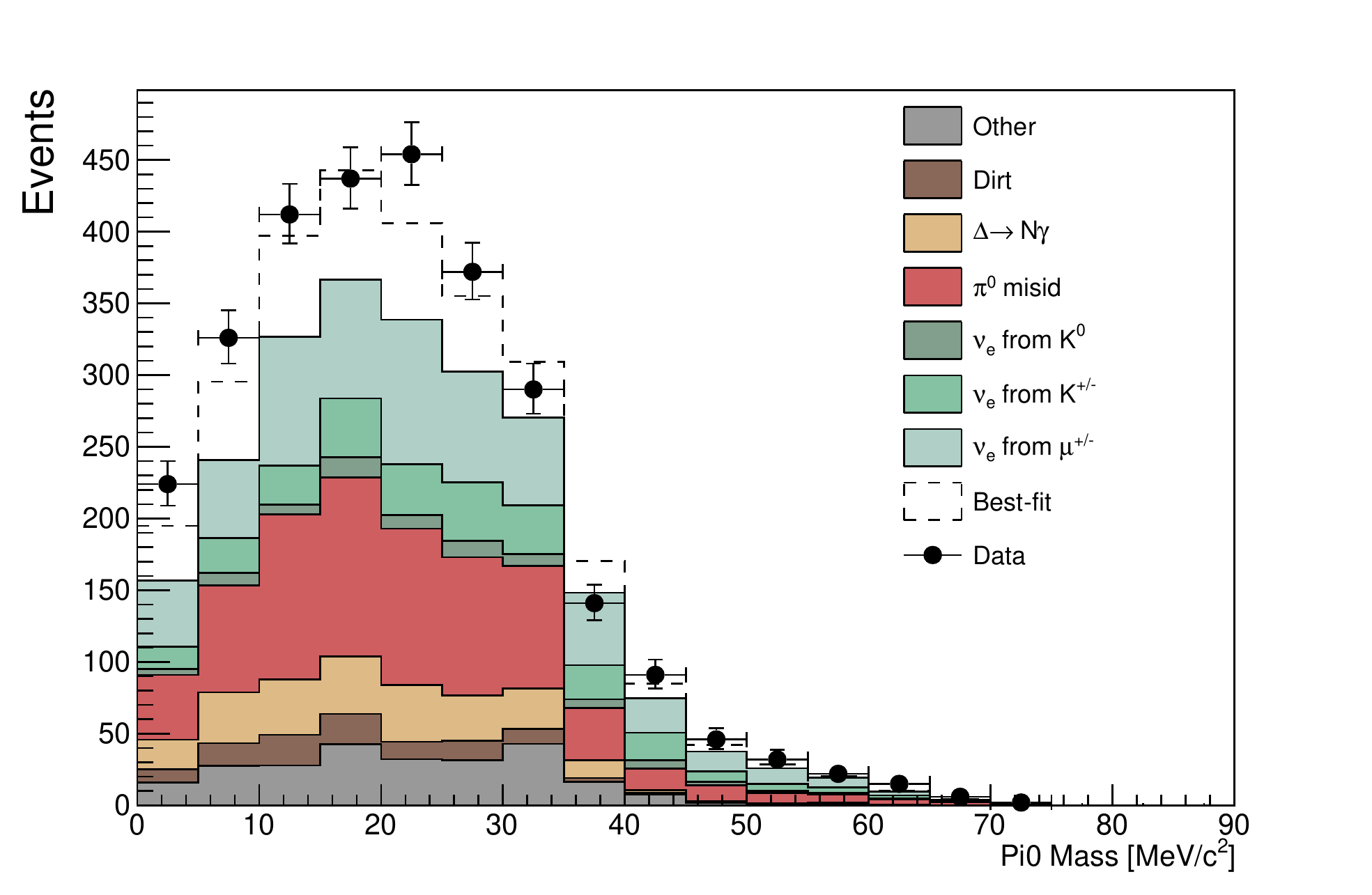}}
\vspace{-0.2in}
\caption{The MiniBooNE neutrino mode
two-ring invariant-mass particle identification distributions, corresponding to the total $18.75 \times 10^{20}$ POT data
in the $200<E_\nu^{QE}<1250$ MeV energy range,
for ${\nu}_e$ CCQE data (points with statistical errors) and background (colored histogram).
The dashed histogram shows the best fit to the neutrino-mode data assuming two-neutrino oscillations.}
\label{PIDpi0}
\vspace{0.1in}
\end{figure}

\begin{figure}[tbp]
\vspace{+0.1in}
\centerline{\includegraphics[angle=0, width=9.0cm]{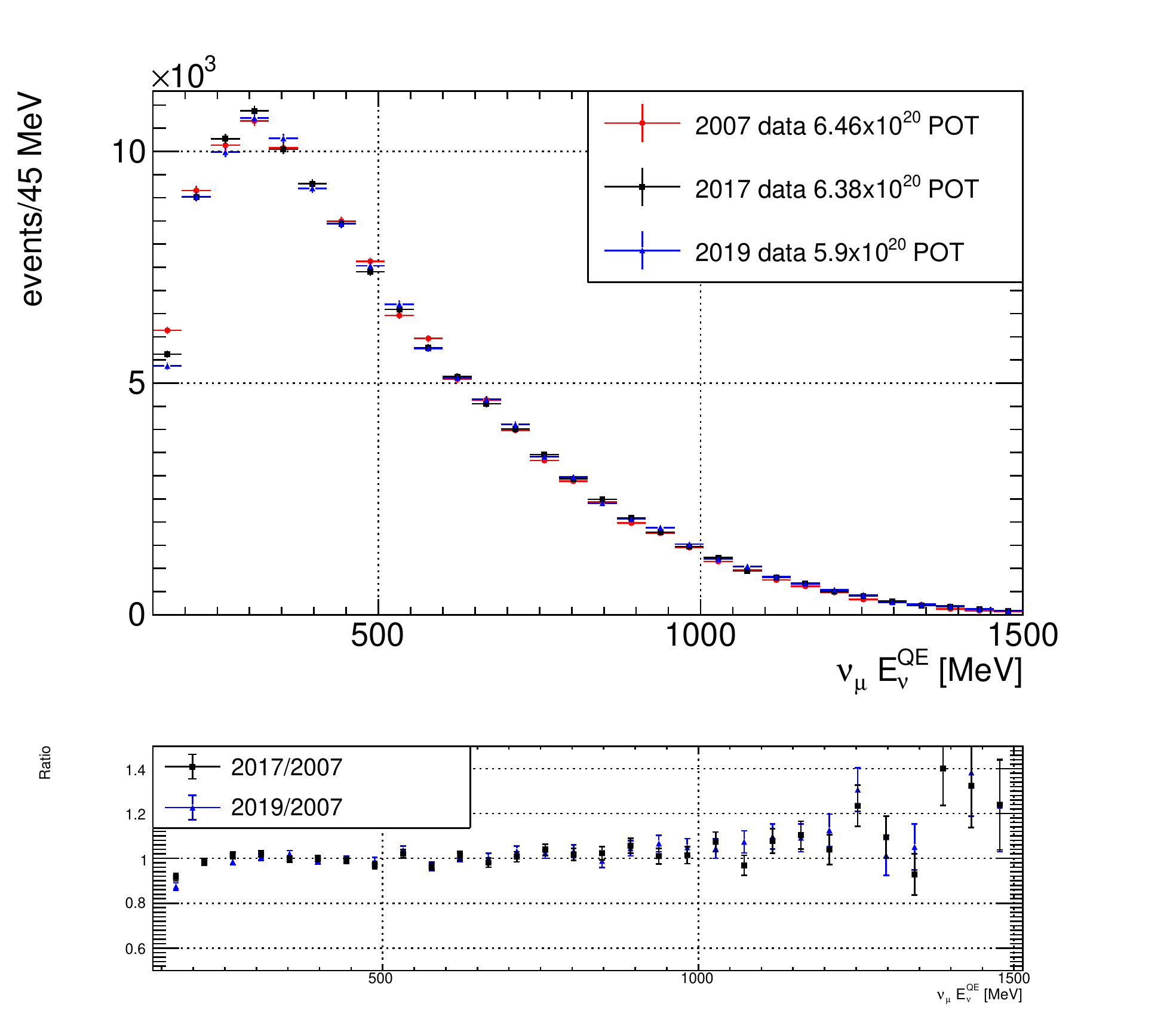}}
\vspace{-0.2in}
\caption{The $\nu_\mu$ CCQE muon visible energy distribution 
for the first, second, and third running periods in neutrino mode.
The events are normalized to the first running period. The bottom plot shows ratios of
the second and third running periods to the first running period.}
\label{numuCCE}
\vspace{0.1in}
\end{figure}

\begin{figure}[tbp]
\vspace{+0.1in}
\centerline{\includegraphics[angle=0, width=9.0cm]{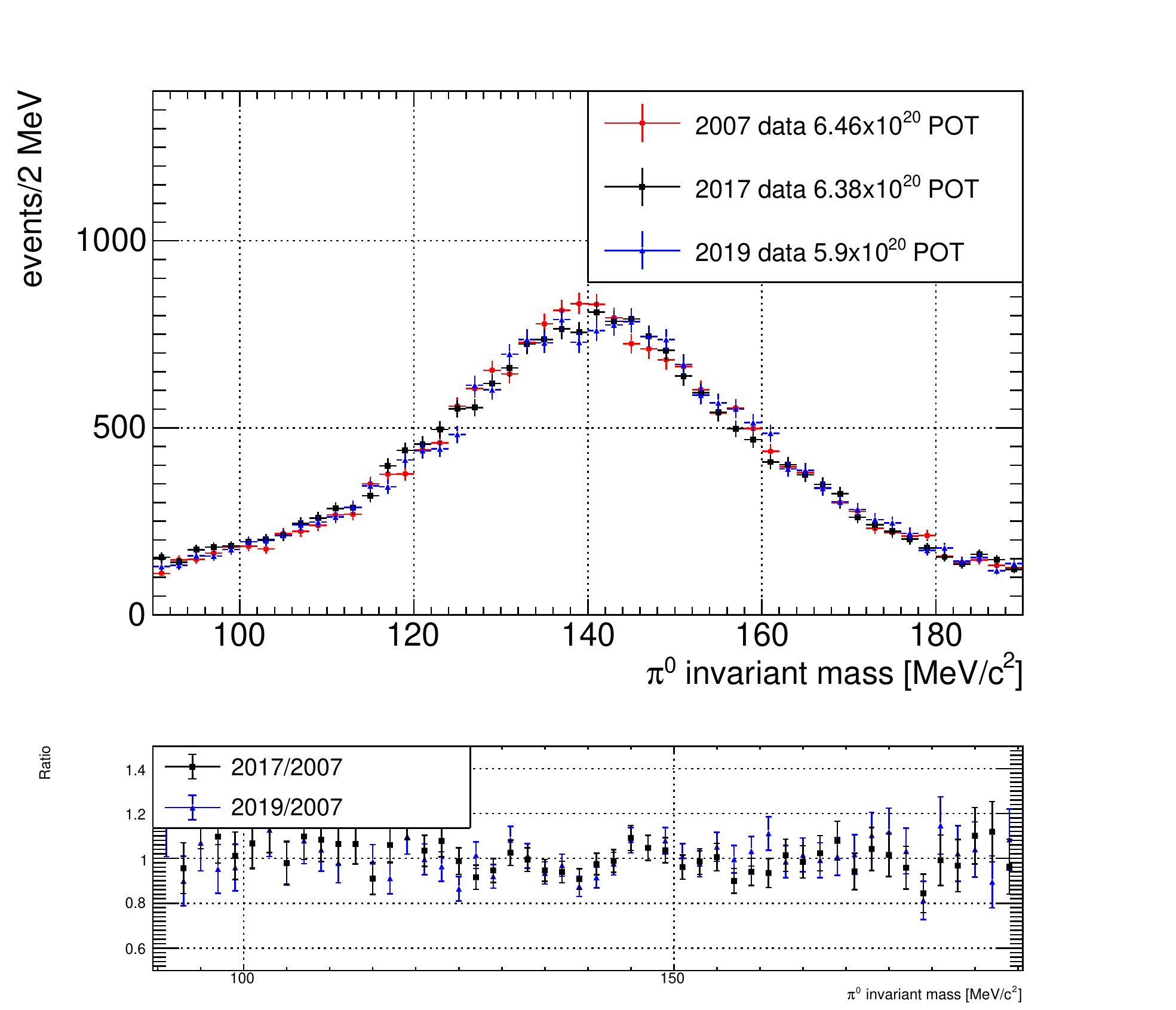}}
\vspace{-0.2in}
\caption{The NC $\pi^0$ mass distribution for the first, second, and third running 
periods in neutrino mode. The events are normailzed to the first running period. The bottom plot shows ratios of
the second and third running periods to the first running period.}
\label{pi0mass}
\vspace{0.1in}
\end{figure}

The data analysis is optimized to measure $\nu_e$-induced CCQE 
events and reject $\nu_\mu$ induced events, and is identical to the previous
analysis \cite{mb_oscnew}. Figs. \ref{PIDem}, \ref{PIDep}, and \ref{PIDpi0} show the 
${\nu}_e$ CCQE data and background for the three particle identification variables in
neutrino mode in the $200<E_\nu^{QE}<1250$~MeV energy range
for the total $18.75 \times 10^{20}$ POT data. The comparison between data and background for the
full range of particle identification variables was shown in the supplementary material of
a previous publication \cite{mb_oscnst}.
The average selection efficiency is
$\sim 20\%$ ($\sim 0.1\%$) for $\nu_e$-induced CCQE events ($\nu_\mu$-induced background events)
generated over the fiducial volume.
The fraction of CCQE events in antineutrino mode that
are from wrong-sign neutrino events was determined from the angular 
distributions of muons created in CCQE interactions and
by measuring CC single $\pi^+$ events \cite{wrong_sign}.
Table \ref{signal_bkgd} shows the predicted but unconstrained $\nu_e$ and 
$\bar{\nu}_e$ CCQE background events for 
the neutrino energy range $200<E_\nu^{QE}<1250$~MeV for both neutrino and antineutrino modes, where there
are approximately twice as many Monte Carlo events compared to data events. 
Table \ref{signal_bkgd} also shows the total constrained background, where the overall normalization
of the $\nu_e$ intrinsic background is constrained by the $\nu_\mu$ CCQE event sample.
The upper limit of 1250 MeV was chosen by the collaboration before unblinding the data in 2007, while 
the lower limit of 200 MeV was chosen in 2013
\cite{mb_oscnew} because it is the lowest energy for reliably reconstructing the
Cherenkov ring of $\nu_\mu$ CCQE events with a visible energy greater than 140 MeV.
From the given detector resolution estimated from the Michel electron spectrum 
(Fig. \ref{Michel}), there is a negligible amount of migration from events below 200 MeV.
Note that the original lower limit was chosen to be 300 MeV before unblinding the data in 2007. 
During the unblinding procedure, the lower limit was increased to 475 MeV due to the low probability
of the two-neutrino oscillation fit and worries about the single-gamma background. However, careful
studies of the single-gamma background were performed after unblinding and convinced the
collaboration that the single-gamma background was estimated correctly within systematic uncertainties
and agreed with theoretical calculations \cite{hill_zhang}. 
Finally, Table \ref{signal_bkgd} shows the expected number of events corresponding to the LSND best fit
oscillation probability of 0.26\%, assuming large $\Delta m^2$ where the oscillations are washed out.
LSND and MiniBooNE have the same average value of L/E, but MiniBooNE has a larger range of L/E. Therefore, 
the appearance probabilities for LSND and MiniBooNE should not be exactly the same at lower L/E values.
Figs. \ref{numuCCE} and \ref{pi0mass} show the $\nu_\mu$ CCQE $E_\nu^{QE}$ energy distribution and the
NC $\pi^0$  mass distribution in neutrino mode for the first, second, and 
third running periods. As shown in the figures, the three running periods have good agreement.

\begin{table}[t]
\vspace{-0.1in}
\caption{\label{signal_bkgd} \em The expected (unconstrained) number of events
for the $200<E_\nu^{QE}<1250$~MeV neutrino 
energy range from all of the backgrounds in the $\nu_e$ and $\bar{\nu}_e$ 
appearance analysis before using the constraint from the CC $\nu_\mu$ events. 
The ``Other'' backgrounds correspond mostly to neutrino-nucleon and neutrino-electron elastic scattering.
Also shown are the constrained background, as well as
the expected number of events corresponding to the LSND best fit 
oscillation probability of 0.26\%, assuming oscillations at large $\Delta m^2$.
The table shows
the diagonal-element systematic plus statistical uncertainties, which become
substantially reduced in the oscillation fits when correlations
between energy bins and between the $\nu_e$ and $\nu_\mu$ events
are included.}
\small
\begin{ruledtabular}
\begin{tabular}{ccc}
Process&Neutrino Mode&Antineutrino Mode \\
\hline
$\nu_\mu$ \& $\bar \nu_\mu$ CCQE & 107.6 $\pm$ 28.2 & 12.9 $\pm$ 4.3 \\
NC $\pi^0$ & 732.3 $\pm$ 95.5 & 112.3 $\pm$ 11.5 \\
NC $\Delta \rightarrow N \gamma$ & 251.9  $\pm$ 35.2 & 34.7 $\pm$ 5.4 \\
External Events & 109.8 $\pm$ 15.9 & 15.3 $\pm$ 2.8 \\
Other $\nu_\mu$ \& $\bar \nu_\mu$ & 130.8 $\pm$ 33.4 & 22.3 $\pm$ 3.5 \\
\hline
$\nu_e$ \& $\bar \nu_e$ from $\mu^{\pm}$ Decay & 621.1 $\pm$ 146.3 & 91.4 $\pm$ 27.6 \\
$\nu_e$ \& $\bar \nu_e$ from $K^{\pm}$ Decay & 280.7  $\pm$ 61.2 & 51.2 $\pm$ 11.0 \\
$\nu_e$ \& $\bar \nu_e$ from $K^0_L$ Decay & 79.6 $\pm$ 29.9 & 51.4 $\pm$ 18.0 \\
Other $\nu_e$ \& $\bar \nu_e$ & 8.8 $\pm$ 4.7 & 6.7 $\pm$ 6.0 \\
\hline
Unconstrained Bkgd. & $2322.6 \pm 258.3$ & $398.2  \pm 49.7$ \\
Constrained Bkgd. & $2309.4 \pm 119.6$ &  $400.6 \pm 28.5$ \\
\hline
Total Data & 2870 &478 \\
Excess & 560.6 $\pm$ 119.6 & 77.4 $\pm$ 28.5 \\
\hline
0.26\% (LSND) $\nu_\mu \rightarrow \nu_e$ & 676.3 & 100.0 \\
\end{tabular}
\vspace{-0.2in}
\end{ruledtabular}
\normalsize
\end{table}

\bigskip
\bigskip

Systematic uncertainties are determined by considering the predicted
effects on the $\nu_\mu$, $\bar{\nu}_{\mu}$, $\nu_e$, and $\bar{\nu}_e$ CCQE rates 
from variations of model parameters that 
include uncertainties in the neutrino and antineutrino flux estimates, 
uncertainties in neutrino cross sections, uncertainties from nuclear effects, and uncertainties in 
detector modeling and reconstruction. 
A covariance matrix in bins of $E^{QE}_{\nu}$ is constructed 
by considering the variation from each source of systematic uncertainty on the $\nu_e$ and $\bar{\nu}_e$ CCQE signal and background, and the
$\nu_\mu$ and $\bar{\nu}_{\mu}$ CCQE prediction as a function of $E_{\nu}^{QE}$.
This matrix includes correlations between any of the $\nu_e$ and $\bar{\nu}_e$ CCQE signal and background and 
$\nu_\mu$ and $\bar{\nu}_{\mu}$ CCQE samples, and is used in the $\chi^2$ calculation of the oscillation fits.

\section{Electron-neutrino Appearance Results}

\begin{figure}[tbp]
\vspace{+0.1in}
\centerline{\includegraphics[angle=0, width=9.0cm]{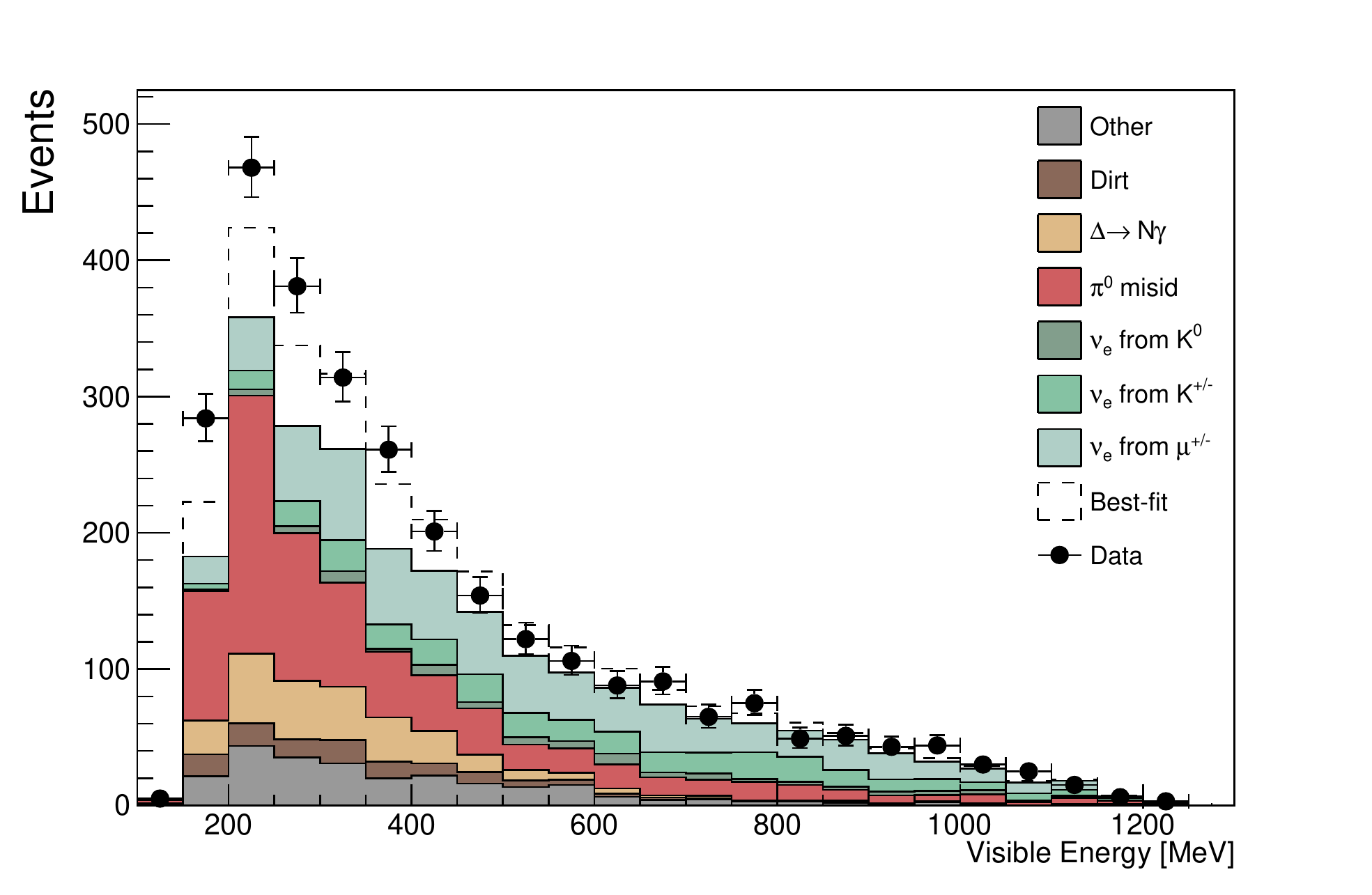}}
\vspace{-0.2in}
\caption{The MiniBooNE neutrino mode
visible energy distributions, corresponding to the total $18.75 \times 10^{20}$ POT data
in the $200<E_\nu^{QE}<1250$ MeV energy range,
for ${\nu}_e$ CCQE data (points with statistical errors) and background (colored histogram).
The dashed histogram shows the best fit to the neutrino-mode data assuming two-neutrino oscillations.}
\label{evis}
\vspace{0.1in}
\end{figure}

\begin{figure}[tbp]
\vspace{+0.1in}
\centerline{\includegraphics[angle=0, width=9.0cm]{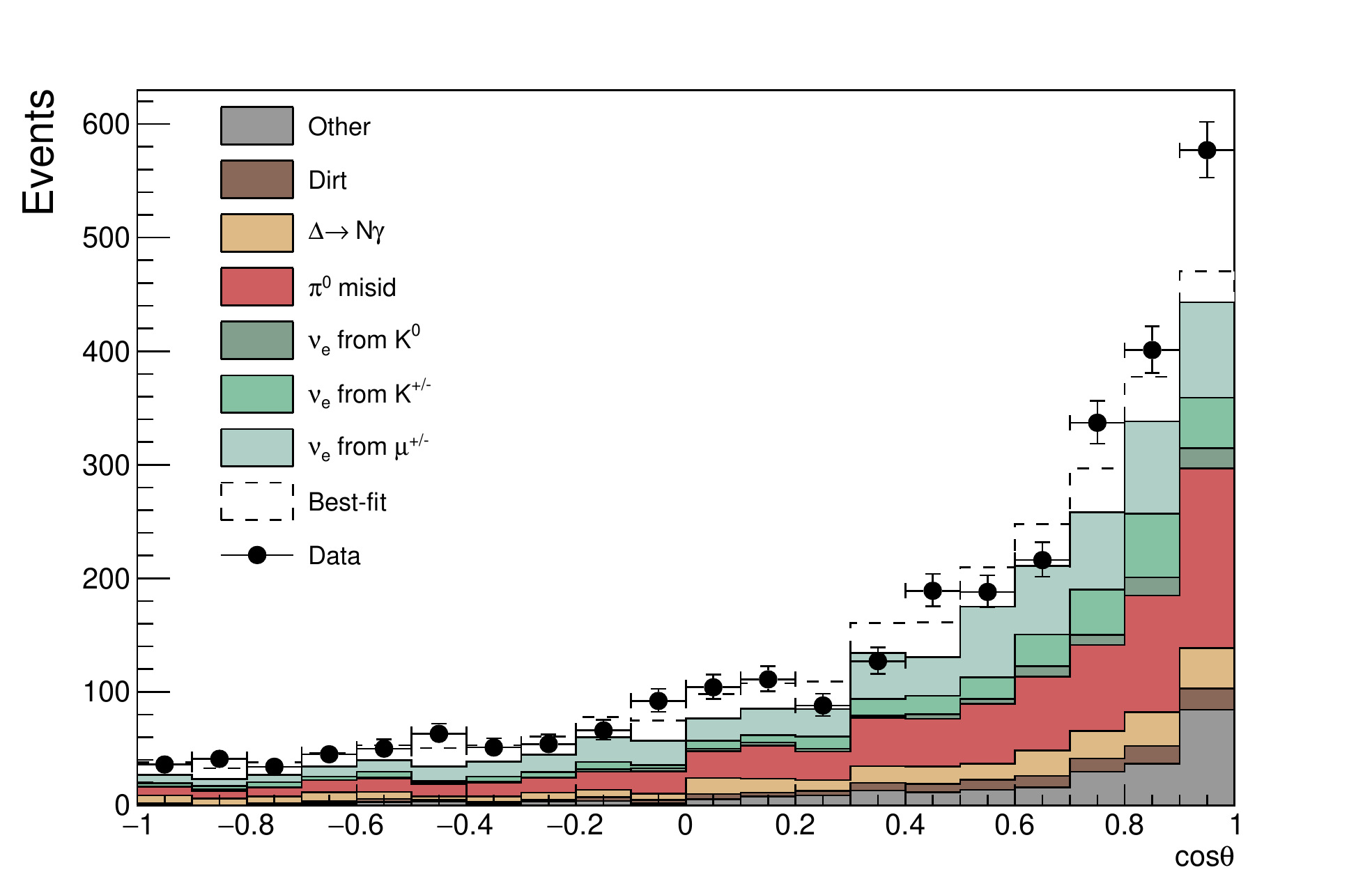}}
\vspace{-0.2in}
\caption{The MiniBooNE neutrino mode
$\cos \theta$ distributions, corresponding to the total $18.75 \times 10^{20}$ POT data
in the $200<E_\nu^{QE}<1250$ MeV energy range,
for ${\nu}_e$ CCQE data (points with statistical errors) and background (colored histogram).
The dashed histogram shows
the best fit to the neutrino-mode data assuming two-neutrino oscillations.}
\label{costh}
\vspace{0.1in}
\end{figure}

\begin{figure}[tbp]
\vspace{+0.1in}
\centerline{\includegraphics[angle=0, width=9.0cm]{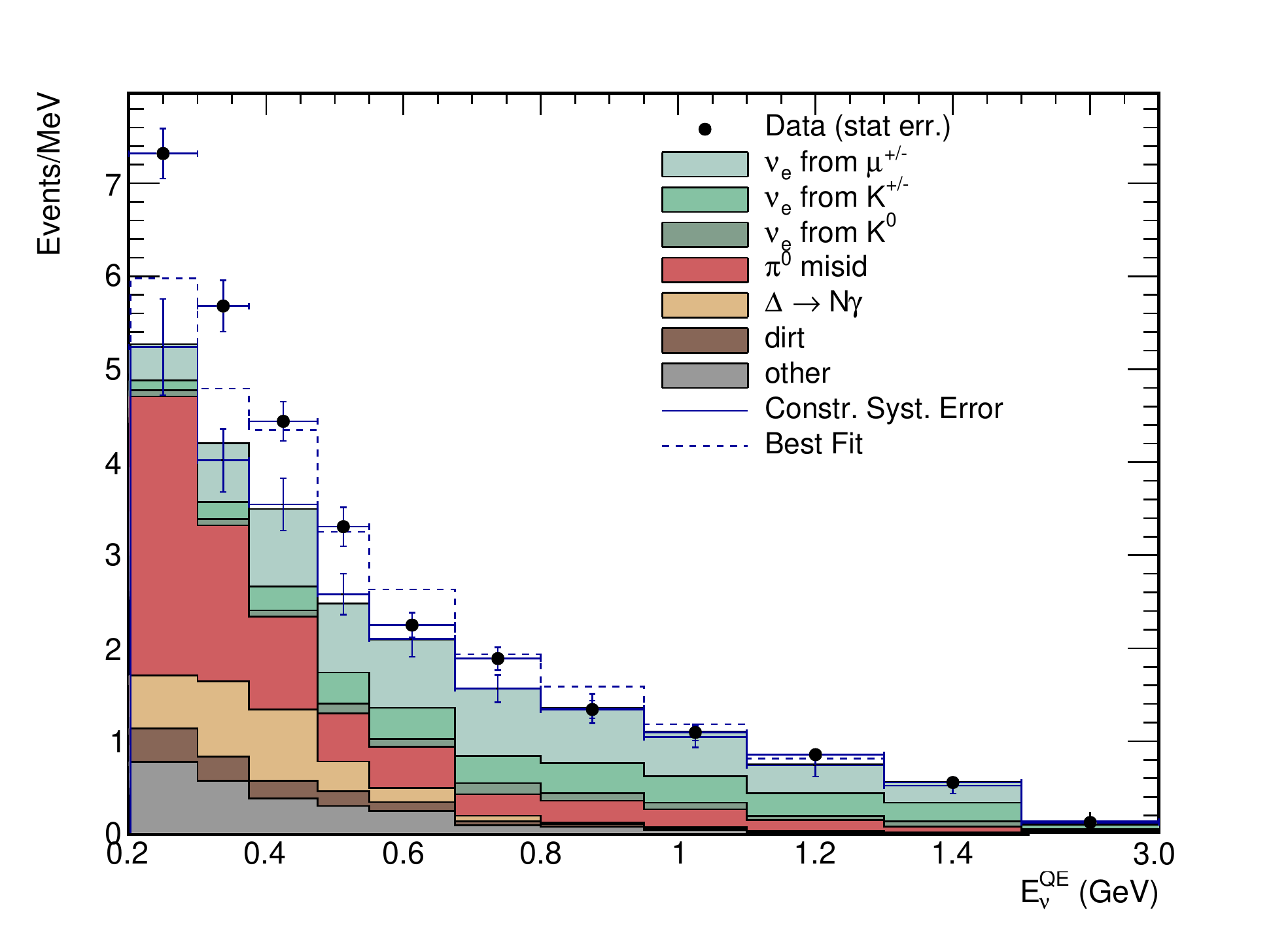}}
\vspace{-0.2in}
\caption{The MiniBooNE neutrino mode 
$E_\nu^{QE}$ distributions, corresponding to the total $18.75 \times 10^{20}$ POT data
in the $200<E_\nu^{QE}<3000$ MeV energy range, 
for ${\nu}_e$ CCQE data (points with statistical errors) and predicted backgrounds (colored histograms). 
The constrained background is shown as additional points with systematic error bars. The dashed histogram 
shows the best fit to the neutrino-mode data assuming two-neutrino oscillations. The last bin is for the 
energy interval from 1500-3000 MeV.}
\label{excessnat}
\vspace{0.1in}
\end{figure}

\begin{figure}[tbp]
\vspace{-0.0in}
\centerline{\includegraphics[angle=0, width=9.0cm]{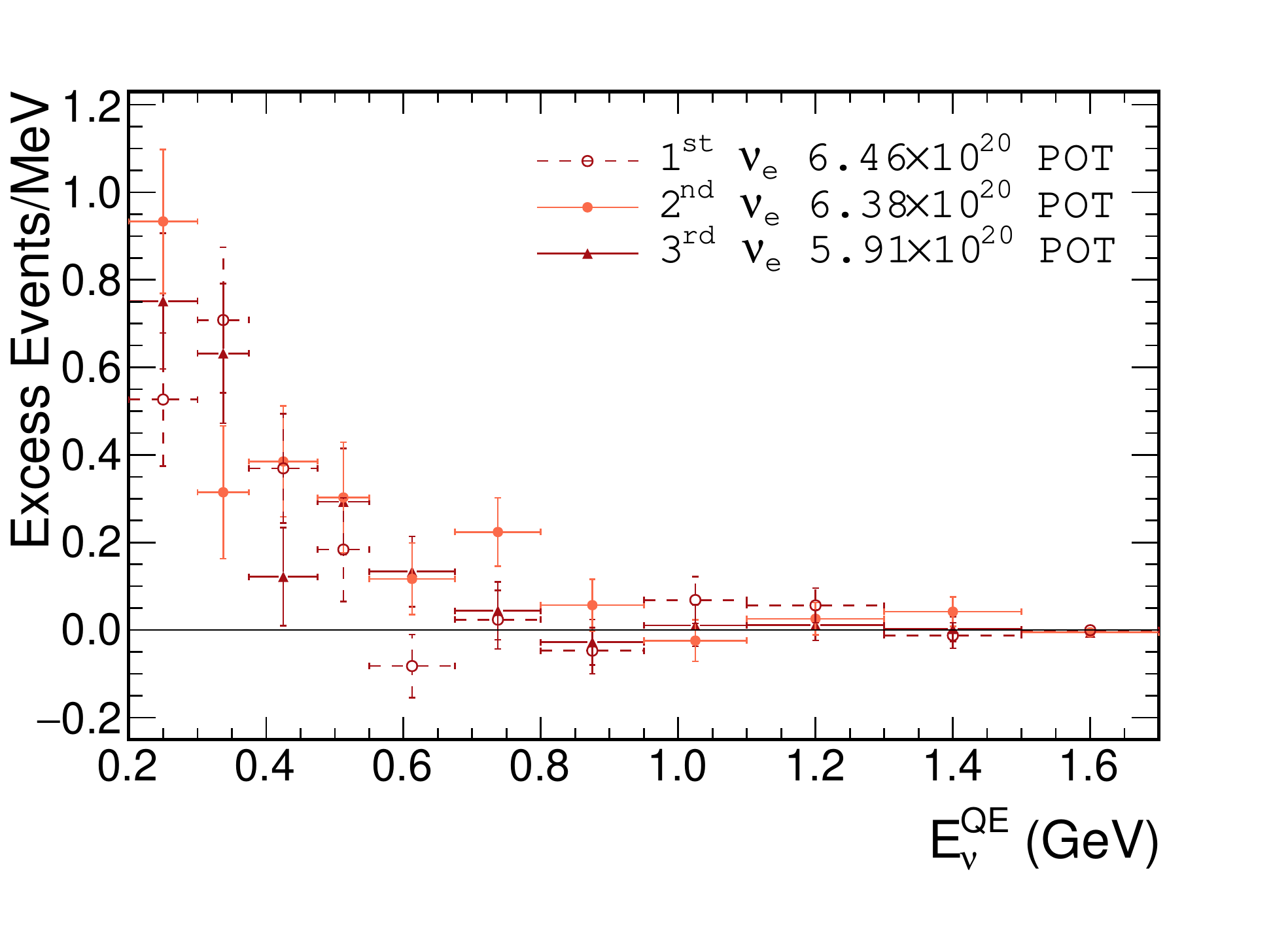}}
\vspace{-0.2in}
\caption{
The total event excesses
in neutrino mode as a function of $E_\nu^{QE}$ for the first, second, and third running periods.
Error bars include only statistical uncertainties.}
\label{Excess_old_new}
\vspace{0.1in}
\end{figure}

Figs. \ref{evis}, \ref{costh}, and \ref{excessnat} show the visible energy, $\cos \theta$, and
$E_\nu^{QE}$ distributions for 
${\nu}_e$ CCQE data and background in
neutrino mode in the $200<E_\nu^{QE}<1250$~MeV energy range
for the total $18.75 \times 10^{20}$ POT data, where $\theta$ is the angle of the reconstructed
electron relative to the incident beam direction.
Each bin of reconstructed $E_\nu^{QE}$
corresponds to a distribution of ``true'' generated neutrino energies,
which can overlap adjacent bins.
In neutrino mode, a total of 2870 data events pass
the $\nu_e$ CCQE event selection requirements with $200<E_\nu^{QE}<1250$~MeV,
compared to a background expectation of $2309.4 \pm 48.1 (stat.) \pm 109.5 (syst.)$ events.
The excess, as shown in Table \ref{excesses}, is then $560.6 \pm 119.6$ events or a $4.7 \sigma$ effect.
Fig. \ref{Excess_old_new} shows the event excesses as a function of $E_\nu^{QE}$ in 
neutrino mode for the first, second, and third running periods.
Combining the MiniBooNE neutrino and antineutrino data \cite{mb_oscnst}, there are a total of 3348 events 
in the $200<E_\nu^{QE}<1250$~MeV energy region, 
compared to a background expectation of $2710.0 \pm 52.1(stat.) \pm 122.2(syst.)$ events. 
This corresponds to a total $\nu_e$ plus $\bar \nu_e$ CCQE excess of $638.0 \pm 52.1(stat.) \pm 122.2(syst.)$ events
with respect to expectation, where the statistical uncertainty is the square root of the background estimate
and the systematic uncertainty includes correlated and uncorrelated systematic parameters. 
The overall significance of the excess, $4.8 \sigma$, is limited by systematic uncertainties,
assumed to be Gaussian, as the statistical significance of
the excess is $12.2 \sigma$.
The fractional unconstrained systematic
uncertainties in the $200<E_\nu^{QE}<1250$~MeV energy range are shown in Table \ref{systs}.

\begin{table}[t]
\vspace{-0.1in}
\caption{\label{excesses} \em The number of data events, background events, and excess events in neutrino
mode for different selection crteria. The errors include both statistical and systematic uncertainties.
Also shown is the significance of each event excess. R is the radius of the reconstructed event interaction
point.
}
\small
\begin{ruledtabular}
\begin{tabular}{ccccc}
Selection&Data&Background&Excess&Significance \\
\hline
$200<E_\nu^{QE}<1250$ MeV \& $R<5m$&2870&$2309.4 \pm 119.6$&$560.6 \pm 119.6$&4.7$\sigma$ \\
$150<E_\nu^{QE}<1250$ MeV \& $R<5m$&3172&$2560.4 \pm 131.5$&$611.6 \pm 131.5$&4.7$\sigma$ \\
$200<E_\nu^{QE}<1250$ MeV \& $R<4m$&1978&$1519.4 \pm 81.9$&$458.6 \pm 81.9$&5.6$\sigma$ \\
$200<E_\nu^{QE}<1250$ MeV \& $R<3m$&864&$673.9 \pm 41.2$&$190.1 \pm 41.2$&4.6$\sigma$ \\
\hline
\end{tabular}
\vspace{-0.2in}
\end{ruledtabular}
\normalsize
\end{table}

\begin{table}[t]
\vspace{-0.1in}
\caption{\label{systs} \em The fractional unconstrained systematic
uncertainties in the $200<E_\nu^{QE}<1250$~MeV energy range.
}
\small
\begin{ruledtabular}
\begin{tabular}{cc}
Systematic Uncertainty&Fraction of Event Excess \\
\hline
Cross Section&35\% \\
Optical Model&23\% \\
$\pi^+$ Production&14\% \\
Neutrino Flux&7\% \\
$K^0$ Production&4\% \\
$K^+$ Production&4\% \\
\hline
\end{tabular}
\vspace{-0.2in}
\end{ruledtabular}
\normalsize
\end{table}

\begin{figure}[tbp]
\vspace{+0.1in}
\centerline{\includegraphics[angle=0, width=9.0cm]{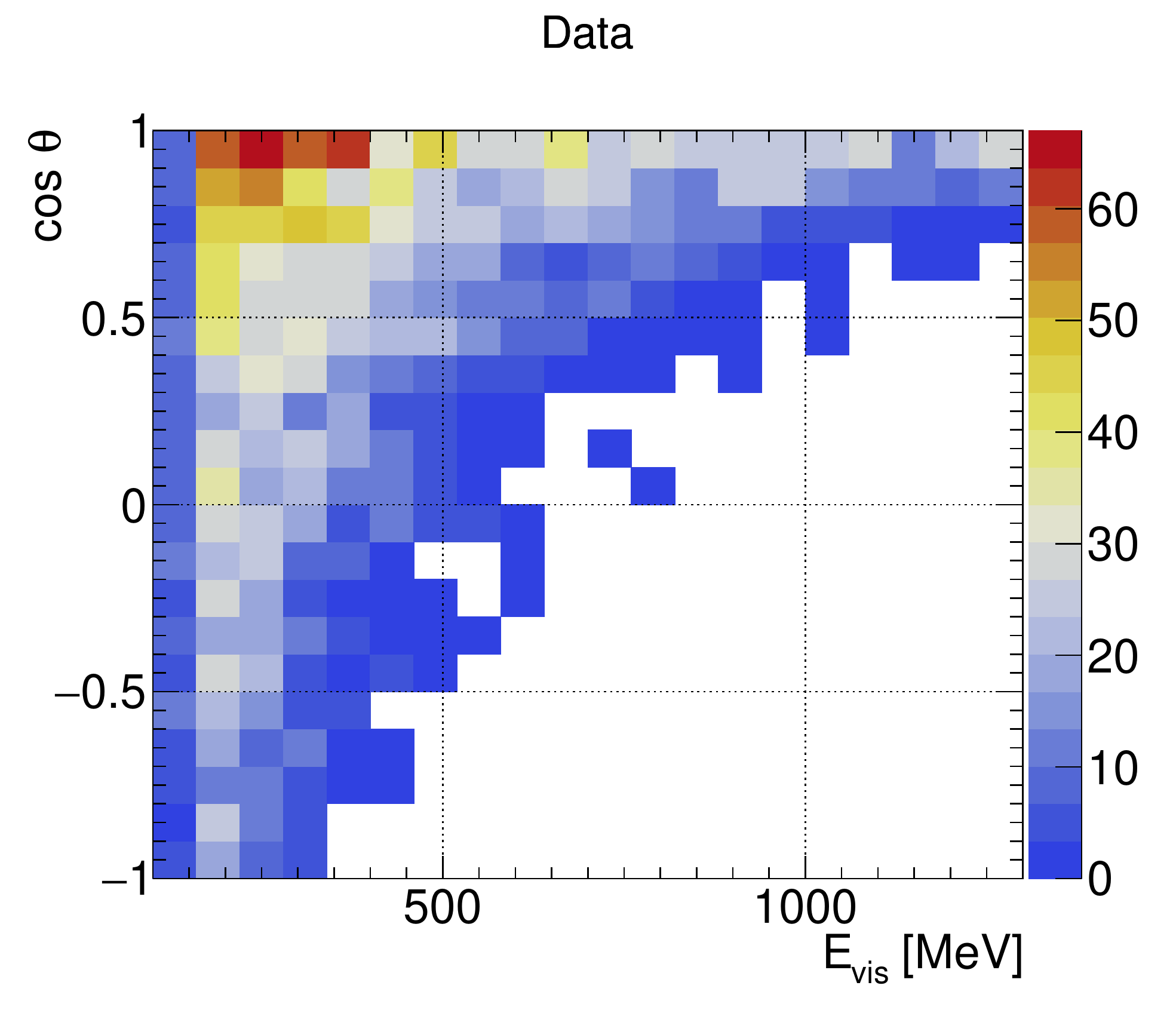}}
\vspace{-0.2in}
\caption{The total numbers of data events in neutrino mode as functions of reconstructed
visible energy and $\cos \theta$.
There are 20 columns of visible energy from 150 to 1250 MeV and 20 rows of $\cos \theta$ from -1
to 1.}
\label{output_D}
\vspace{0.1in}
\end{figure}

\begin{figure}[tbp]
\vspace{+0.1in}
\centerline{\includegraphics[angle=0, width=9.0cm]{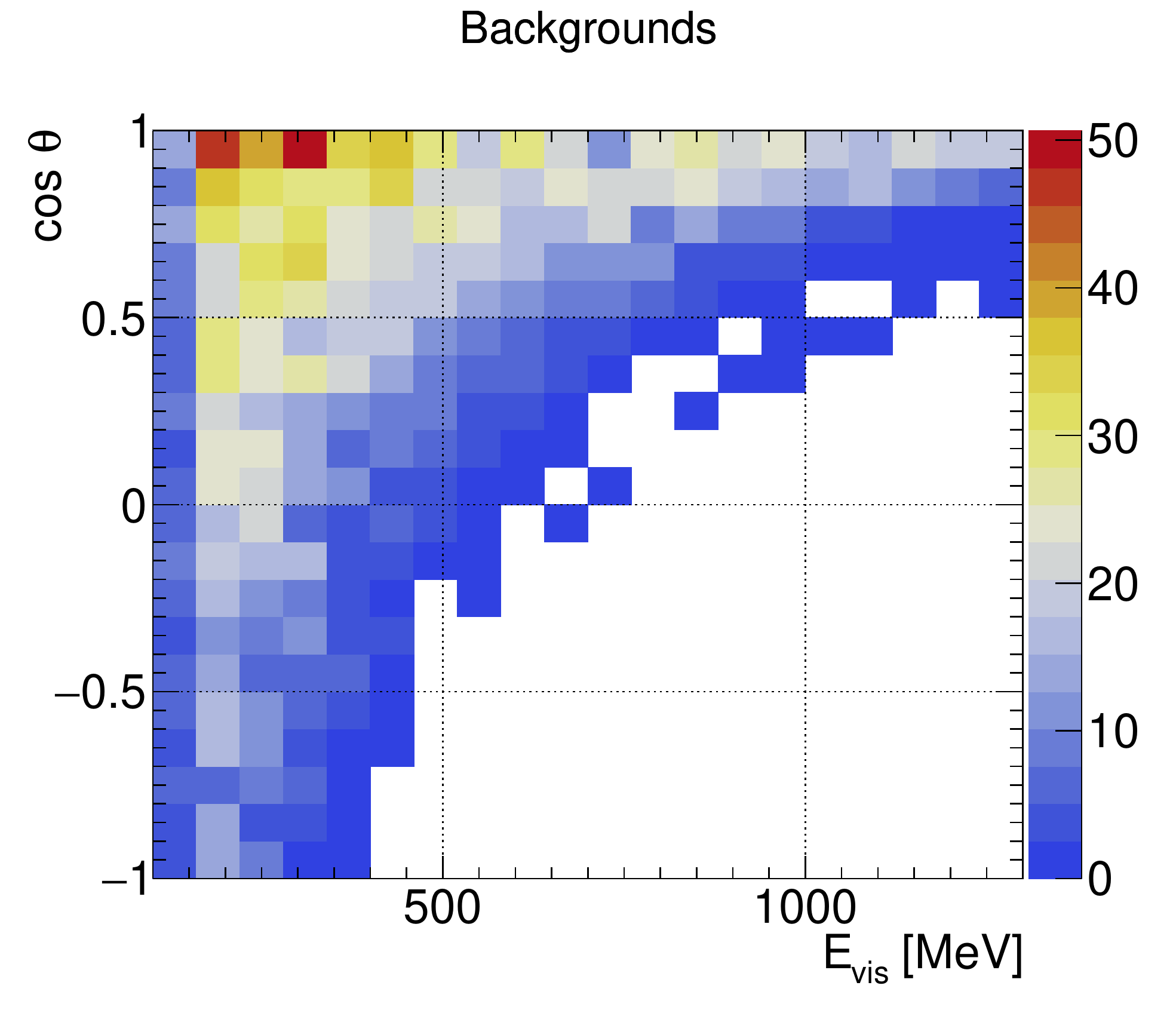}}
\vspace{-0.2in}
\caption{The total numbers of background events in neutrino mode
as functions of reconstructed visible energy and $\cos \theta$.
There are 20 columns of visible energy from 150 to 1250 MeV and 20 rows of $\cos \theta$ from -1
to 1.}
\label{output_B}
\vspace{0.1in}
\end{figure}

\begin{figure}[tbp]
\vspace{+0.1in}
\centerline{\includegraphics[angle=0, width=9.0cm]{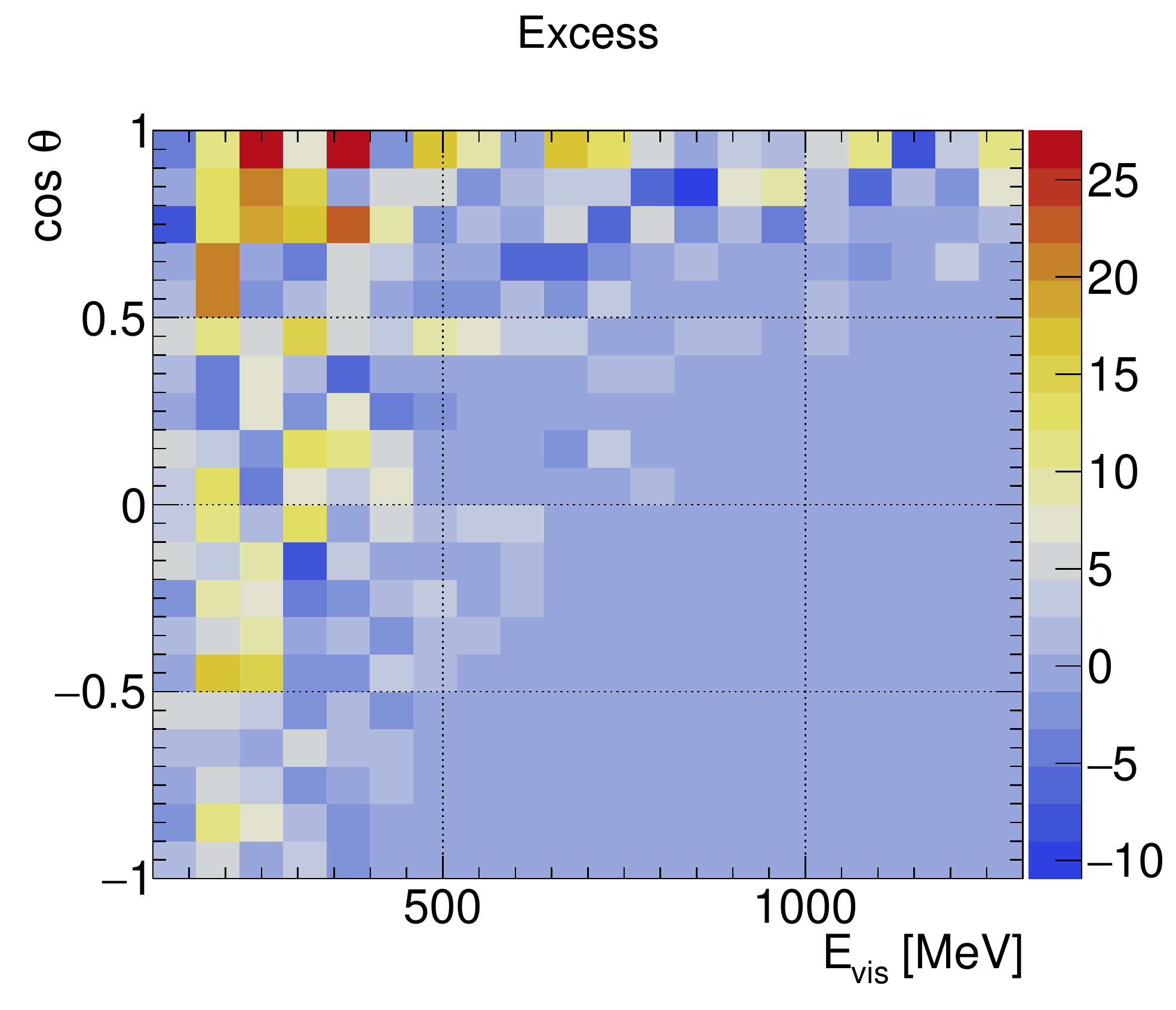}}
\vspace{-0.2in}
\caption{The total numbers of excess events in neutrino mode as functions of reconstructed
visible energy and $\cos \theta$.
There are 20 columns of visible energy from 150 to 1250 MeV and 20 rows of $\cos \theta$ from -1
to 1.}
\label{output_E}
\vspace{0.1in}
\end{figure}

\begin{figure}[tbp]
\vspace{+0.1in}
\centerline{\includegraphics[angle=0, width=18.0cm]{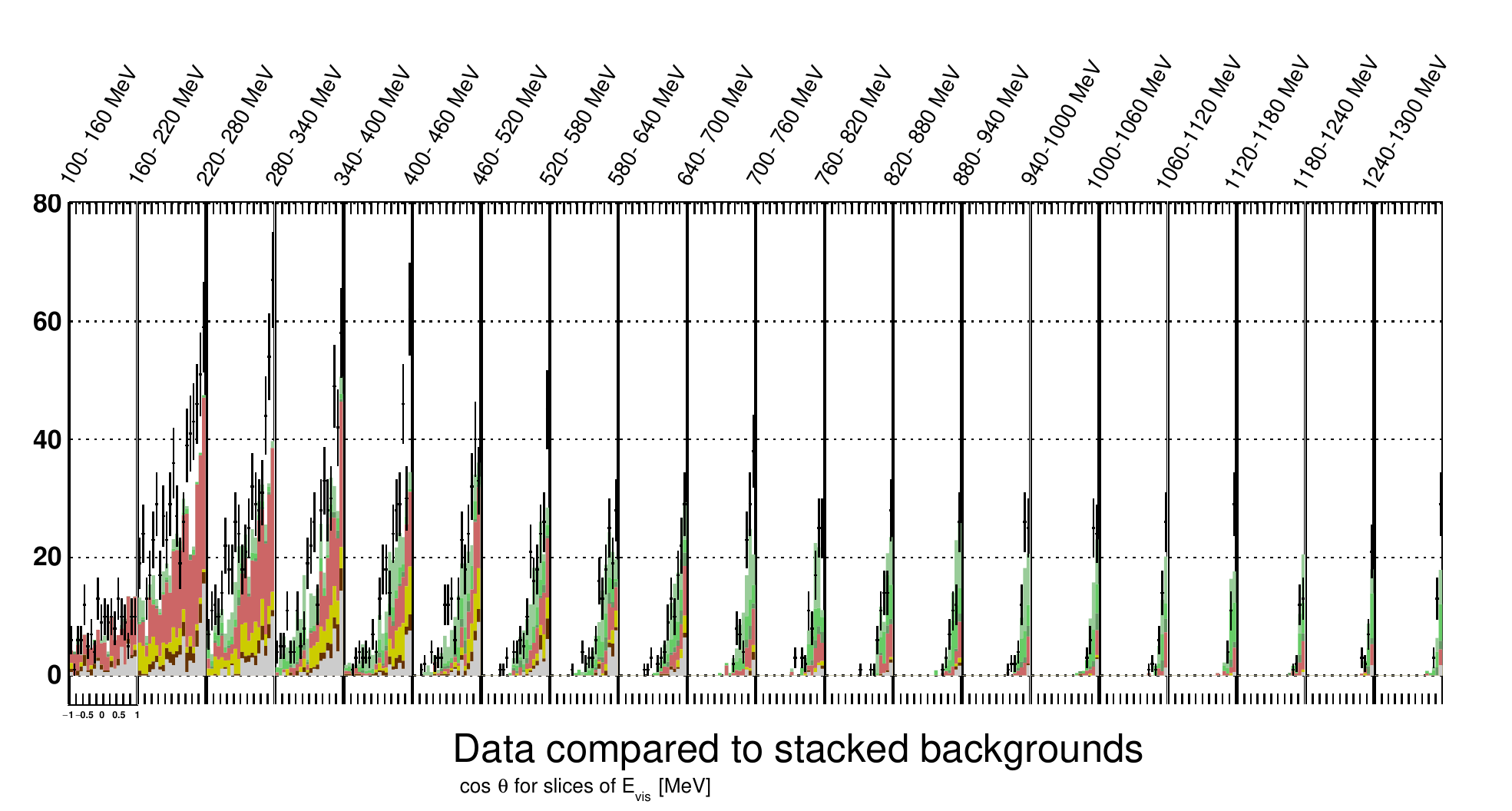}}
\vspace{-0.2in}
\caption{The $\cos \theta$ distribution of data events (points with error bars)
and background events (colored histogram) in neutrino mode for the 20 different visible energy bins
from 150 to 1250 MeV.}
\label{1D_data}
\vspace{0.1in}
\end{figure}

\begin{figure}[tbp]
\vspace{+0.1in}
\centerline{\includegraphics[angle=0, width=18.0cm]{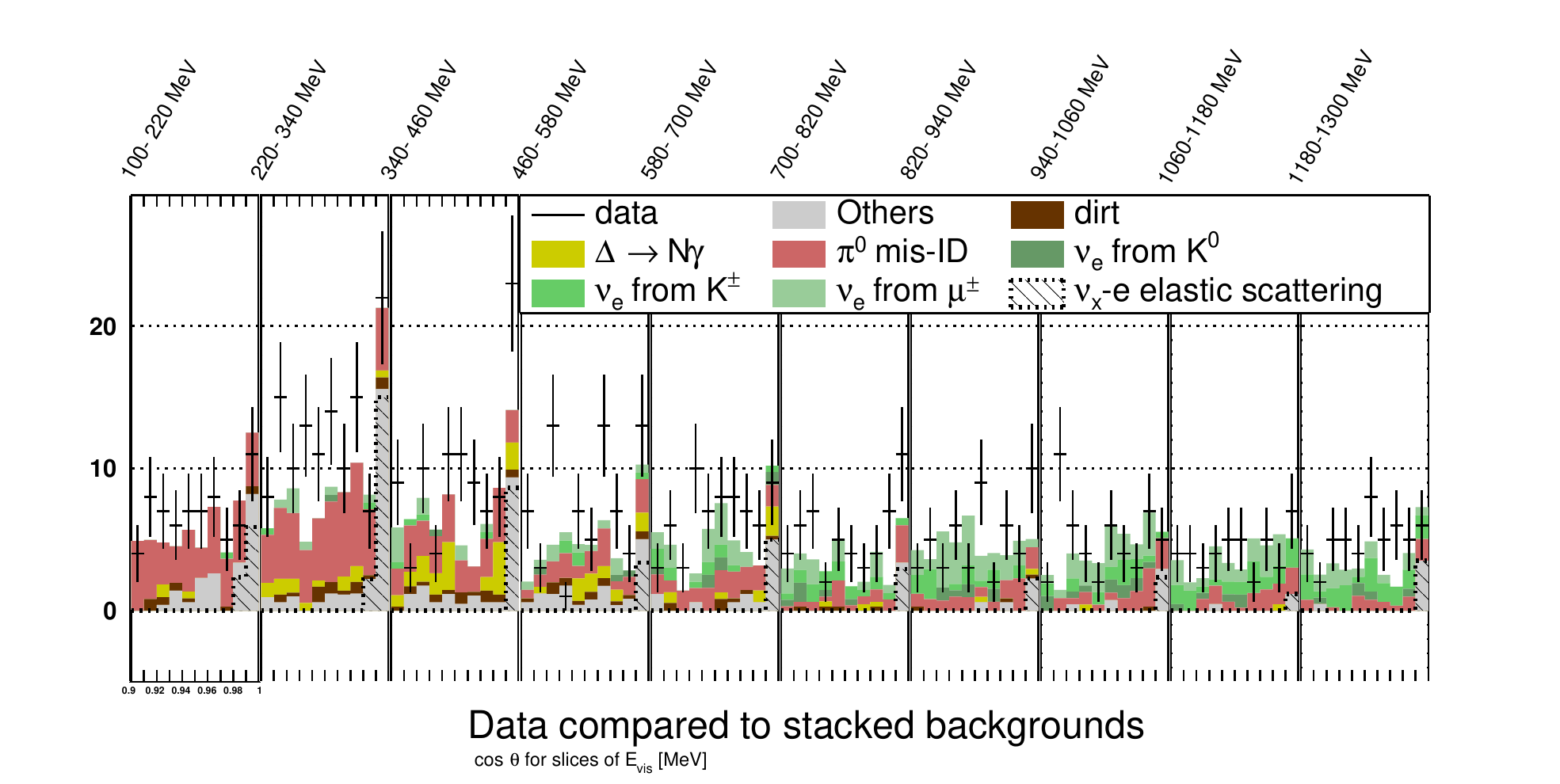}}
\vspace{-0.2in}
\caption{The $\cos \theta$ distributions from 0.9 to 1 of data events (points with statistical errors)
and background events (histogram) in neutrino mode for 10 different visible energy bins
from 150 to 1250 MeV. Neutrino-electron elastic scattering events are shown as the hatched 
region in the ``Others'' category.}
\label{Uz_E}
\vspace{0.1in}
\end{figure}

In order to test physics models, the numbers of data events, unconstrained background events, 
and excess events in neutrino mode
with visible energy between 150 and 1250 MeV are shown
in Figs. \ref{output_D}, \ref{output_B}, and \ref{output_E} as functions of visible energy and $\cos \theta$.
In these figures, there are 20 columns of visible energy from 150 to 1250 MeV and 20 rows of $\cos \theta$ from -1
to 1. There are a total of 3182 data events, 2568.8 background events and 613.2 excess events. Fig. \ref{1D_data}
shows the $\cos \theta$ distribution of data and background events
for the 20 different energy bins, while Fig. \ref{Uz_E} shows the $\cos \theta$ distributions from 0.9 to 1
for 10 different visible energy bins. Neutrino-electron elastic
scattering events are shown as the hatched region in the ``Others'' category.

\begin{figure}[tbp]
\vspace{+0.1in}
\centerline{\includegraphics[angle=0, width=9.0cm]{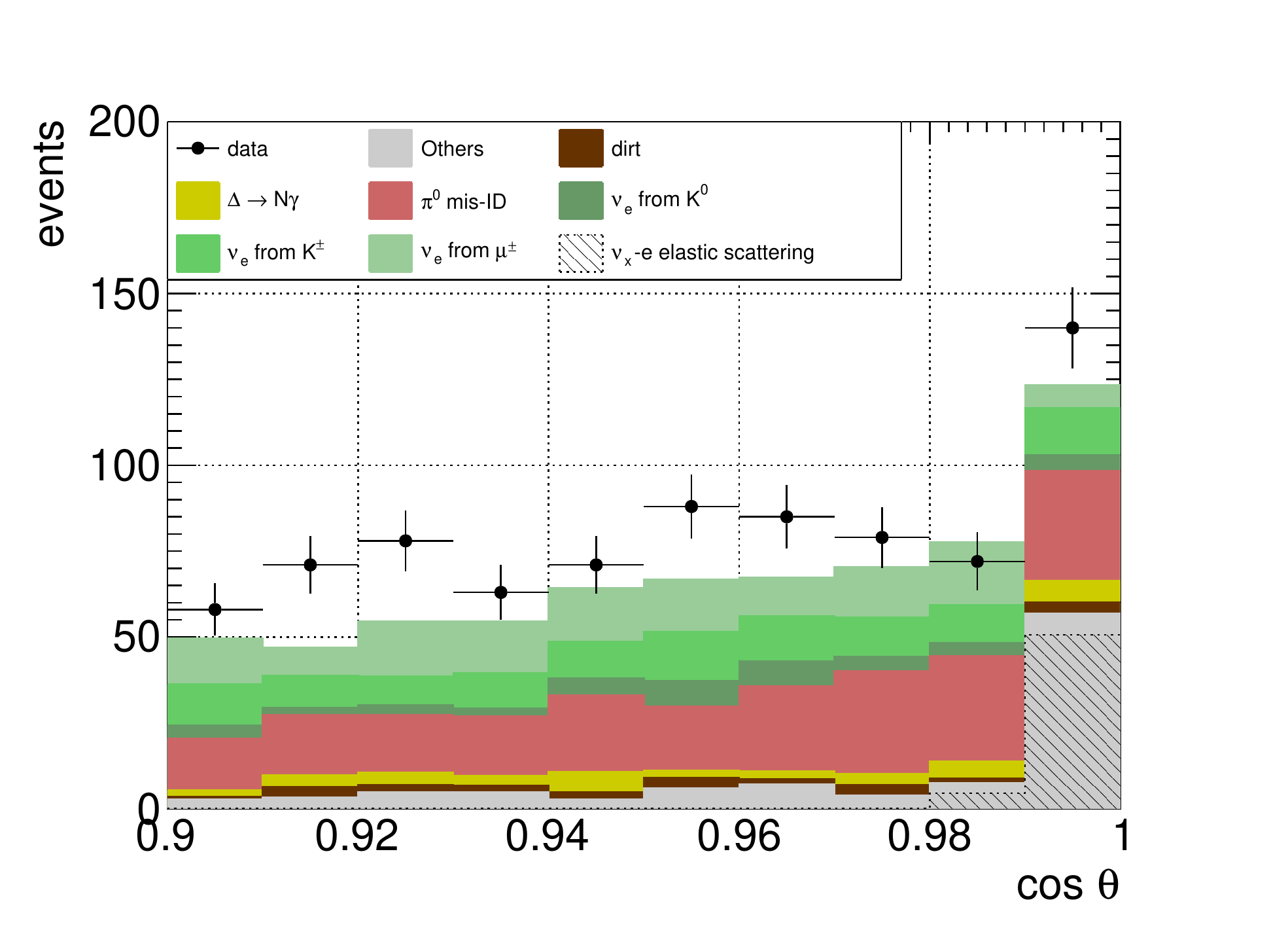}}
\vspace{-0.2in}
\caption{The MiniBooNE neutrino mode
$\cos \theta$ distribution for $\cos \theta > 0.9$, corresponding to the total $18.75 \times 10^{20}$ POT neutrino data
in the visible energy range from 150 to 1250 MeV,
for ${\nu}_e$ CCQE data (points with statistical errors) and predicted backgrounds (colored histograms).
Neutrino-electron elastic scattering events are shown as the hatched region in the ``Others'' category.}
\label{Uz0.9}
\vspace{0.1in}
\end{figure}

Fig. \ref{Uz0.9} shows the number of data and background
events as a function of $\cos \theta$ for $\cos \theta > 0.9$, where neutrino-electron elastic
scattering events are shown as the hatched region in the ``Others'' category and contribute to the 
$\cos \theta > 0.98$ bins. The neutrino-electron elastic events constitute 53\% (89\%) of the
``Others'' category for $\cos \theta > 0.90$ ($\cos \theta > 0.99$), and the category also includes 
neutrino-nucleon charged-current and neutral-current scattering events.
As shown in the figure, the excess of data events over background events
is approximately the same in each bin.

\begin{figure}[tbp]
\vspace{+0.1in}
\centerline{\includegraphics[angle=0, width=9.0cm]{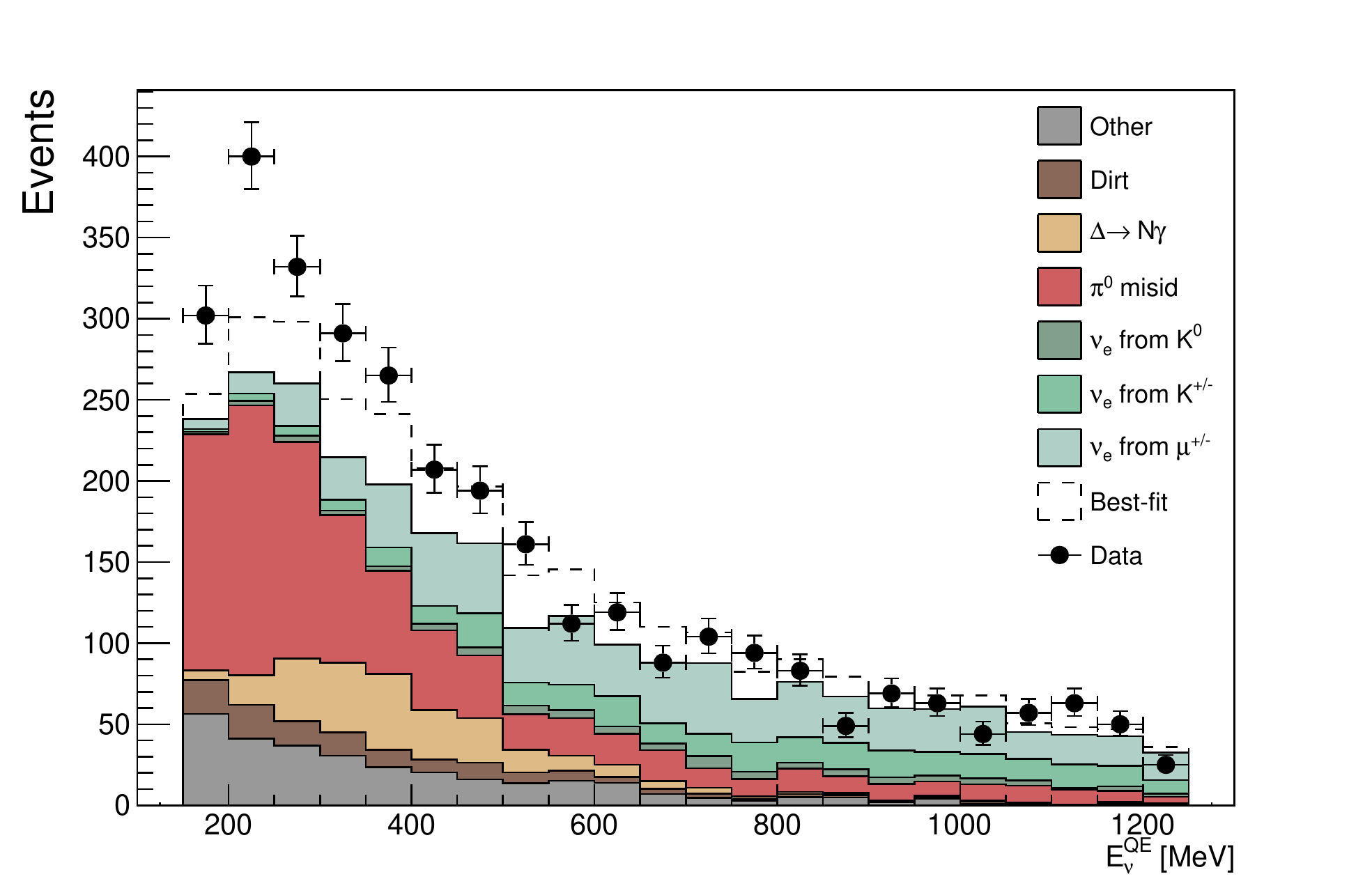}}
\vspace{-0.2in}
\caption{The MiniBooNE neutrino mode
$E_\nu^{QE}$ distributions, corresponding to the total $18.75 \times 10^{20}$ POT neutrino data
in the $150<E_\nu^{QE}<1250$ MeV energy range,
for ${\nu}_e$ CCQE data (points with statistical errors) and predicted backgrounds (colored histograms).
The dashed histogram
shows the best fit to the neutrino-mode data assuming two-neutrino oscillations.} 
\label{EnuQE150}
\vspace{0.1in}
\end{figure}

\begin{figure}[tbp]
\vspace{+0.1in}
\centerline{\includegraphics[angle=0, width=9.0cm]{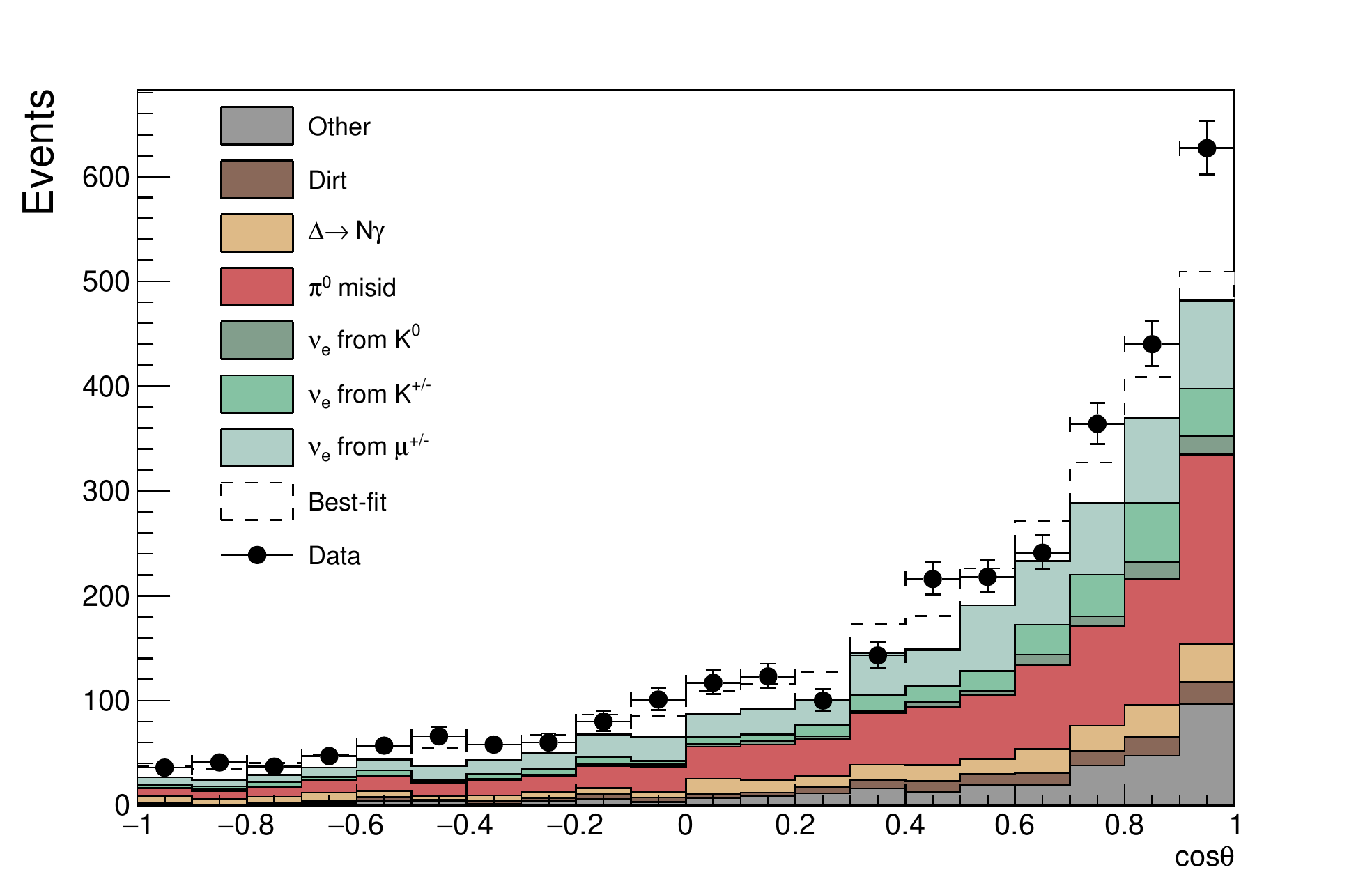}}
\vspace{-0.2in}
\caption{The MiniBooNE neutrino mode
$\cos \theta$ distributions, corresponding to the total $18.75 \times 10^{20}$ POT neutrino data
in the $150<E_\nu^{QE}<1250$ MeV energy range,
for ${\nu}_e$ CCQE data (points with statistical errors) and predicted backgrounds (colored histograms).
The dashed histogram
shows the best fit to the neutrino-mode data assuming two-neutrino oscillations.} 
\label{Uz150}
\vspace{0.1in}
\end{figure}

\begin{figure}[tbp]
\vspace{-0.0in}
\centerline{\includegraphics[angle=0, width=18.0cm]{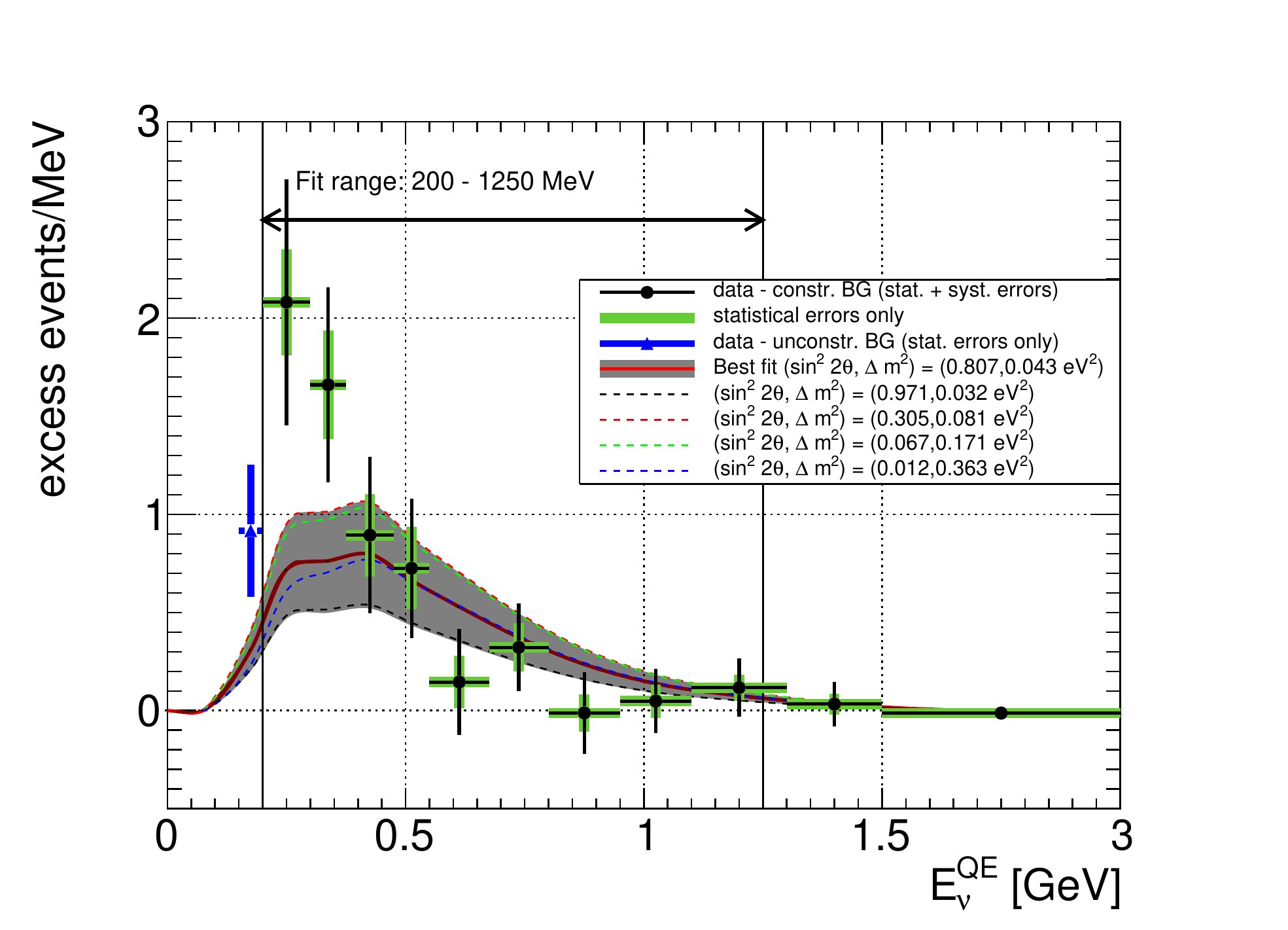}}
\vspace{-0.2in}
\caption{
The total event excess 
in neutrino mode, corresponding to $18.75 \times 10^{20}$ POT in the $150<E_\nu^{QE}<3000$ MeV energy range. 
The solid curve shows the best fit to the neutrino-mode and antineutrino-mode data
in the $200 < E_\nu^{QE} < 1250$ MeV energy range
assuming two-neutrino oscillations, while the dashed curves show selected points on the 1 sigma contour.
Also shown is the 1-sigma allowed band.
The outer error bars include statistical plus constrained systematic uncertainties, while the inner 
error bars show the statistical uncertainties. The lowest energy data point
shows only the unconstrained background statistical uncertainty.}
\label{Excess_new}
\vspace{0.1in}
\end{figure}

Figs. \ref{EnuQE150} and \ref{Uz150} show the $E_\nu^{QE}$ and $\cos \theta$ distributions 
for the $150<E_\nu^{QE}<1250$ MeV energy range, and the total event excess as a function of $E_\nu^{QE}$ is shown in 
Fig. \ref{Excess_new}. 
The solid curve on the latter plot shows the two-neutrino oscillation prediction at the
best-fit point ($\sin^22\theta = 0.807$, $\Delta m^2=0.043$ eV$^2$). 
The lowest energy data point has less acceptance than the other data points due to the requirement 
that the visible energy be greater than 140 MeV.
Table \ref{excesses} lists the number of data events, background events, excess events, and
excess significance for the $150<E_\nu^{QE}<1250$ MeV energy range.

\section{Neutrino Oscillation Fits}

\begin{figure}[tbp]
\vspace{-0.25in}
 \centerline{\includegraphics[width=9.0cm]{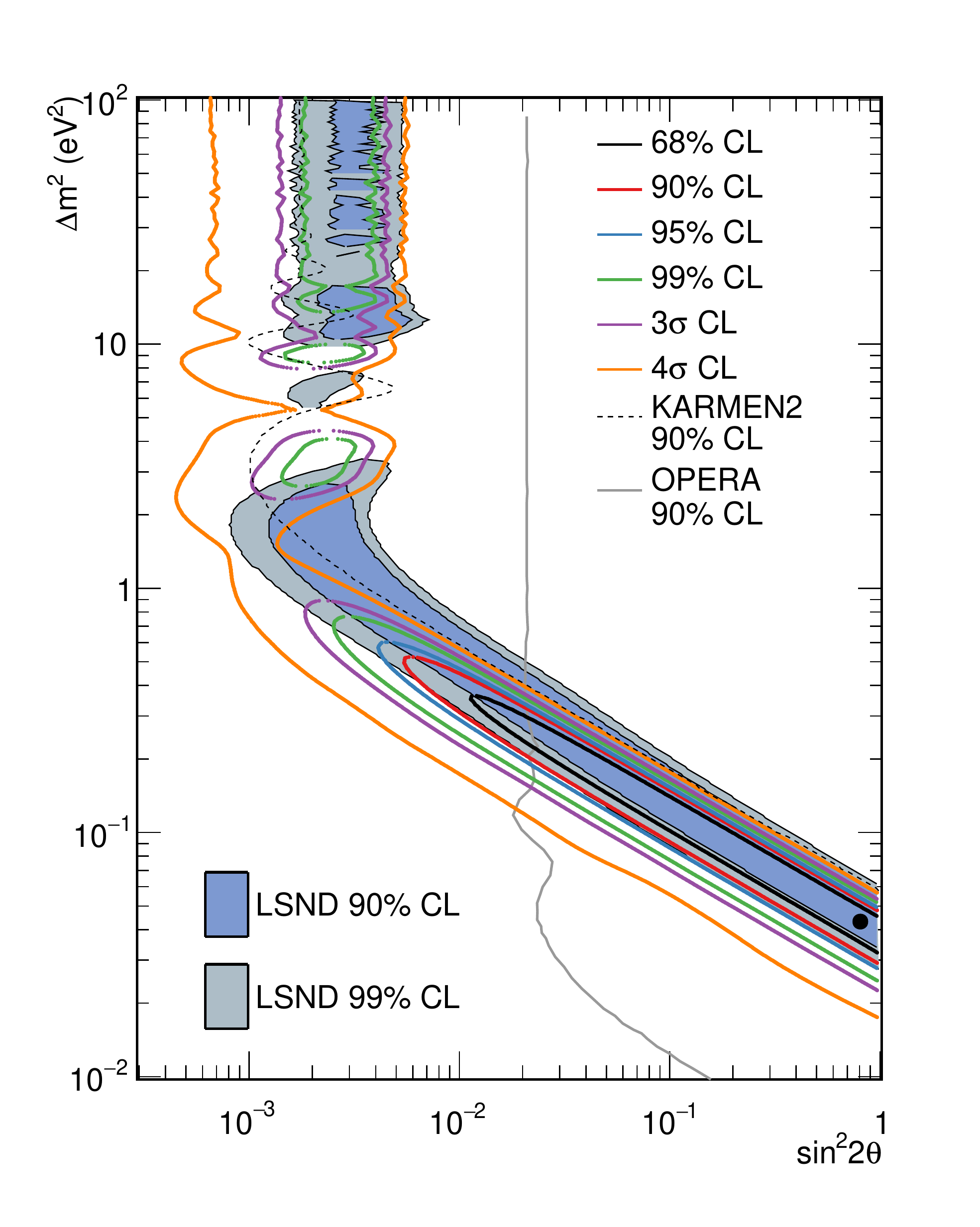}}
 \vspace{-0.3in}
\caption{MiniBooNE allowed regions for combined neutrino mode ($18.75
  \times 10^{20}$ POT) and antineutrino mode ($11.27 \times 10^{20}$ POT)
data sets for events with
$200 < E^{QE}_{\nu} < 3000$ MeV within a two-neutrino oscillation model.
The shaded areas show the 90\% and 99\% C.L. LSND
$\bar{\nu}_{\mu}\rightarrow\bar{\nu}_e$ allowed
regions. The black point shows the MiniBooNE best fit point. Also shown are 90\% C.L. limits
from the KARMEN \cite{karmen}
and OPERA \cite{opera} experiments.}
\label{limitab2}
\vspace{0.1in}
\end{figure}

Fig.~\ref{limitab2} shows the MiniBooNE allowed regions in both neutrino mode and antineutrino
mode \cite{mb_oscnst}
for events with $200 < E^{QE}_{\nu} < 3000$ MeV within a two-neutrino oscillation model.
For this oscillation fit the entire data set is used and includes the $18.75 \times 10^{20}$ POT
data in neutrino mode and the $11.27 \times 10^{20}$ POT data in antineutrino mode.
Also shown are 90\% C.L. limits from the KARMEN \cite{karmen}
and OPERA \cite{opera} experiments.
The best combined neutrino oscillation fit occurs at
($\sin^22\theta$, $\Delta m^2$) $=$ (0.807, 0.043 eV$^2$).
The $\chi^2/ndf$ for the best-fit point in the energy range $200<E_\nu^{QE}<1250$~MeV
is 21.7/15.5 with a probability of $12.3\%$, and the background-only fit
has a  $\chi^2$ probability of $3 \times 10^{-7}$ relative to the best oscillation fit
and a $\chi^2/ndf = 50.7/17.3$ with a probability of $0.01\%$.

\begin{figure}[tbp]
\vspace{-0.0in}
\centerline{\includegraphics[angle=0, width=9.0cm]{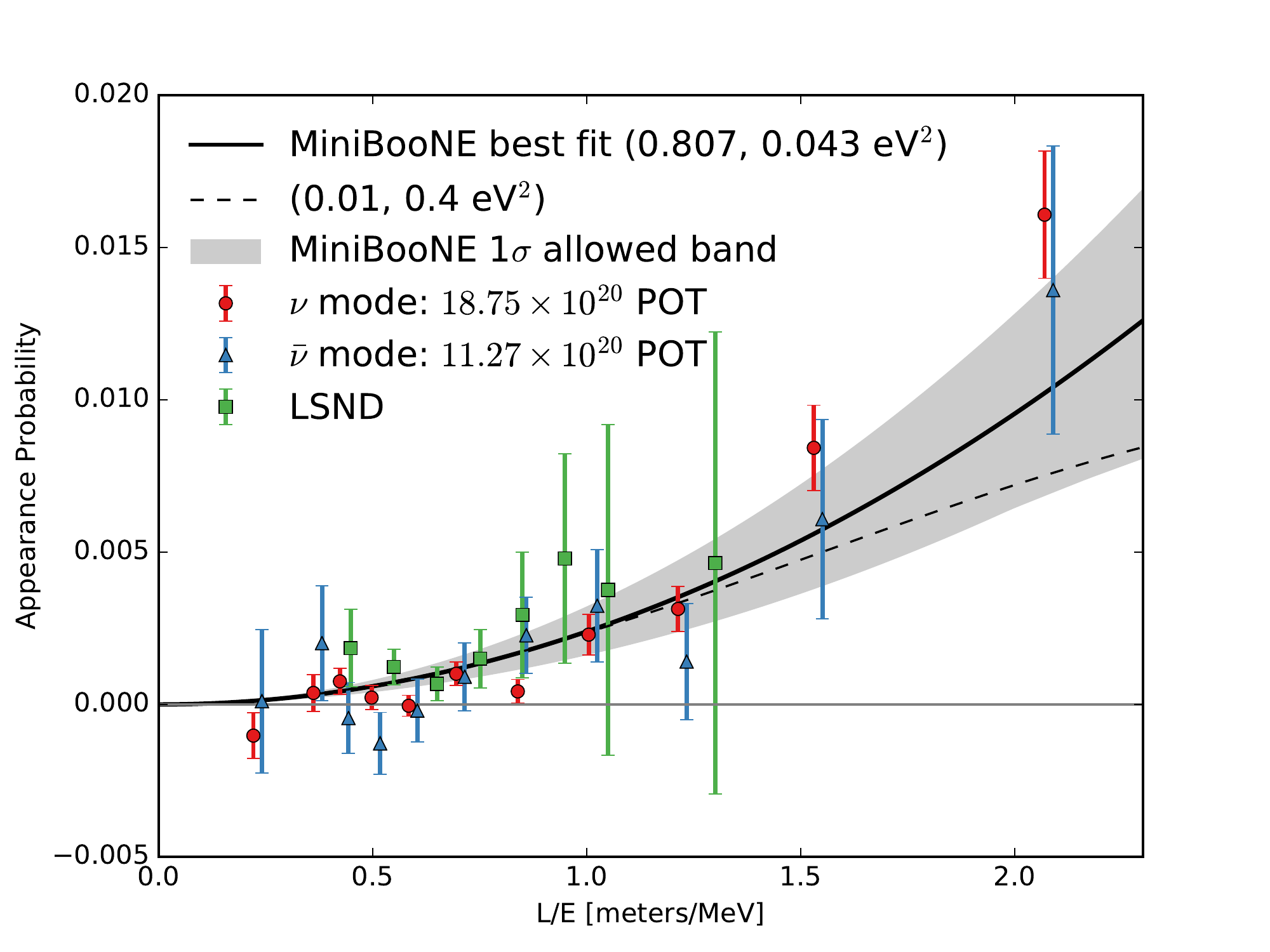}}
\vspace{-0.2in}
\caption{A comparison between the $L/E_\nu^{QE}$
distributions for the MiniBooNE data excesses in neutrino mode
($18.75 \times 10^{20}$ POT) and antineutrino mode ($11.27
\times 10^{20}$ POT) to the $L/E$ distribution from LSND \cite{lsnd}.
The error bars show
statistical uncertainties only. The curves show fits to the
MiniBooNE data, assuming two-neutrino oscillations, while the shaded area is the MiniBooNE $1 \sigma$
allowed band.
The best-fit curve corresponds to ($\sin^22\theta$, $\Delta m^2$) $=$ (0.807, 0.043 eV$^2$), while 
the dashed curve corresponds to a $1 \sigma$ fit point
at ($\sin^22\theta$, $\Delta m^2$) $=$ (0.01, 0.4 eV$^2$).}
\label{LoverE}
\vspace{0.1in}
\end{figure}

Fig. \ref{LoverE} compares the $L/E_\nu^{QE}$
distributions for the MiniBooNE data excesses in neutrino mode and
antineutrino mode to the $L/E$ distribution from LSND \cite{lsnd}.
The error bars show statistical uncertainties only. As shown in the figure,
there is agreement among all three data sets.
Assuming two-neutrino oscillations, the curves show fits to the
MiniBooNE data described above.
The significance of the combined LSND ($3.8 \sigma$) \cite{lsnd}
and MiniBooNE ($4.8 \sigma$) excesses is $6.1 \sigma$, which is obtained by adding the significances
in quadrature, as the two experiments have completely different neutrino energies, neutrino fluxes,
reconstructions, backgrounds, and systematic uncertainties.

\section{Background Studies and Constraints}

Constraints have been placed on the various backgrounds in Table \ref{signal_bkgd} by direct
measurements of these backgrounds in the MiniBooNE detector.
The $\nu_\mu$ CC background has been well measured \cite{mb_numuccqe}
by using the Michel electrons from muon decay to identify
the event topology. Likewise, the NC $\pi^0$ background has also been well measured \cite{mb_pi0}
by reconstructing the
two-gamma invariant mass. 

\begin{figure}[tbp]
\vspace{-0.25in}
 \centerline{\includegraphics[angle=0,width=9.0cm]{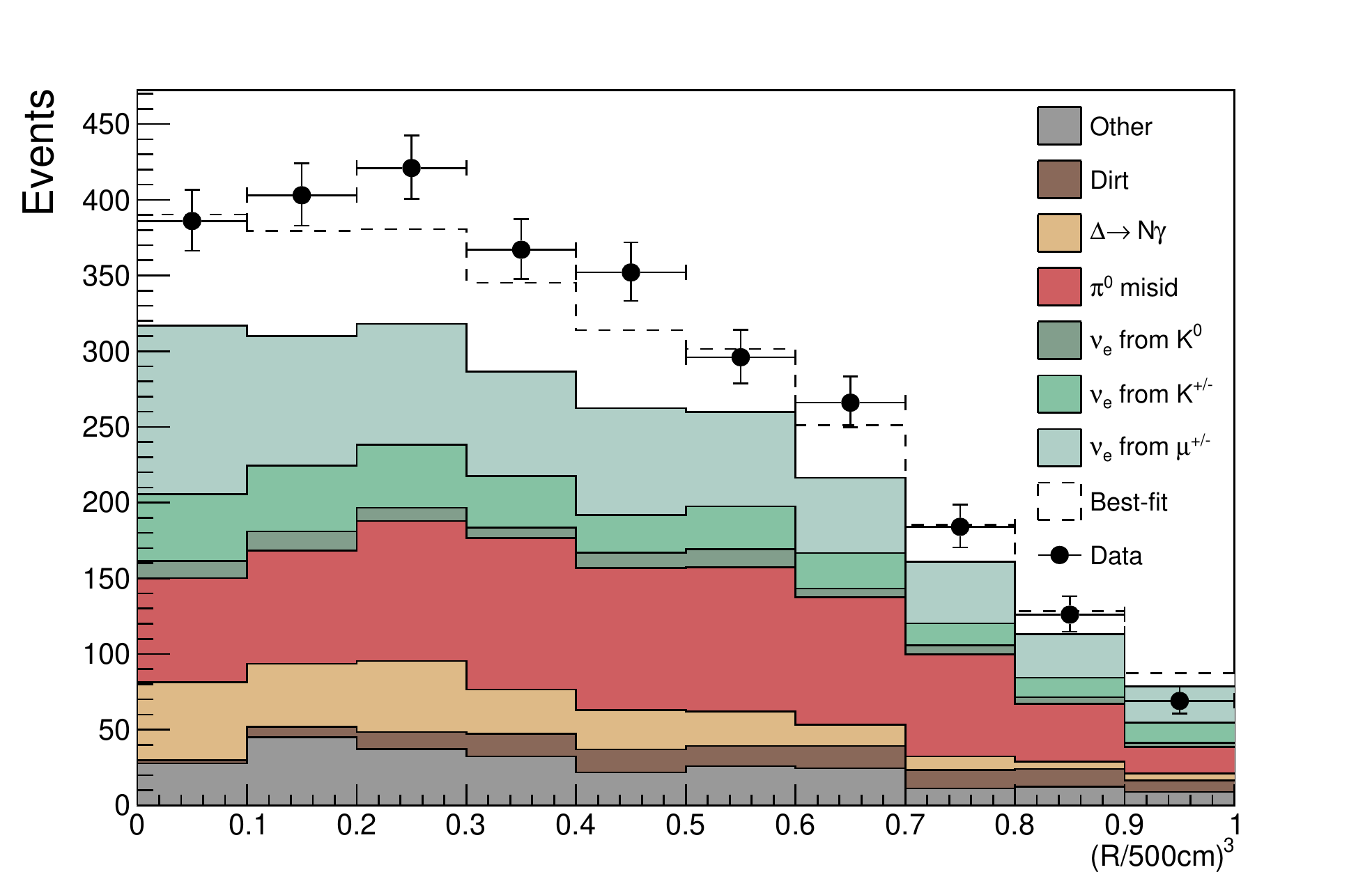}}
 \vspace{-0.3in}
\caption{The MiniBooNE radial vertex distribution, corresponding to the total $18.75 \times 10^{20}$ 
POT data in neutrino mode in the $200<E_\nu^{QE}<1250$ MeV energy range, 
for ${\nu}_e$ CCQE data (points with statistical errors) and background (histogram).
The dashed histogram shows the best fit to the neutrino-mode data assuming two-neutrino oscillations.}
\label{Rdist}
\vspace{0.1in}
\end{figure}

\begin{figure}[tbp]
\vspace{-0.25in}
 \centerline{\includegraphics[angle=0,width=18.0cm]{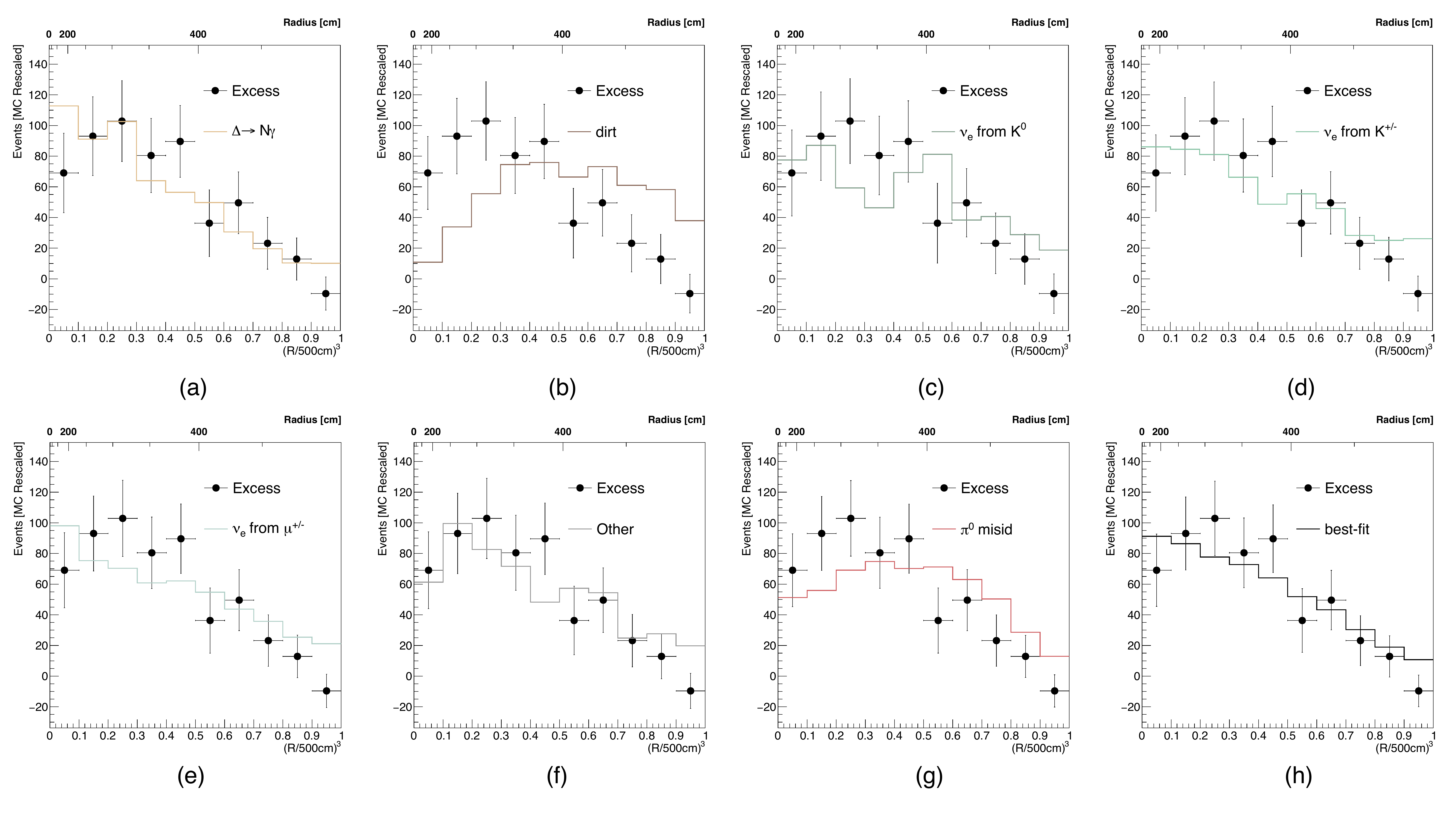}}
 \vspace{-0.3in}
\caption{The excess event radial distributions in neutrino mode with only statistical errors
in the $200<E_\nu^{QE}<1250$ MeV energy range, where
different processes are normalized to explain the event excess. The different processes are the
following: (a) $\Delta \rightarrow N \gamma$; (b) External Events; (c) $\nu_e$ \& $\bar \nu_e$ from $K^0_L$ Decay;
(d) $\nu_e$ \& $\bar \nu_e$ from $K^{\pm}$ Decay; (e) $\nu_e$ \& $\bar \nu_e$ from $\mu^{\pm}$ Decay; 
(f) Other $\nu_e$ \& $\bar \nu_e$; (g) NC $\pi^0$; (h) Best Fit oscillations.}
\label{RExcess}
\vspace{0.1in}
\end{figure}

In addition, a fit to the vertex radial distribution, shown in 
Fig. \ref{Rdist}, allows a 
constraint to be placed on the NC $\pi^0$ background, due to this background having more events near the edge of the
5 m radius fiducial volume. (NC $\pi^0$ events near the edge of the fiducial volume have a greater chance of one
photon leaving the detector with the remaining photon then mis-reconstructing as an electron candidate.) 
Fig. \ref{RExcess} shows the excess event radial distributions, where
different processes are normalized to explain the event excess, while  
Table \ref{Rfit} shows the result of log-likelihood shape-only fits to the radial
distribution and the multiplicative factor that is required for each hypothesis 
to explain the observed event excess. The two-neutrino oscillation hypothesis
fits the radial distribution best with a $\chi^2 = 8.4/9 ndf$, while the NC $\pi^0$ hypothesis has a worse fit
with a $\chi^2 = 17.2/9 ndf$. 
The intrinsic $\nu_e$ backgrounds have a worse $\chi^2$ than the two-neutrino 
oscillation hypothesis due to higher energy $\nu_e$ events having a different radial distribution than lower
energy $\nu_e$ events.

\begin{figure}[tbp]
\vspace{-0.25in}
 \centerline{\includegraphics[angle=0,width=9.0cm]{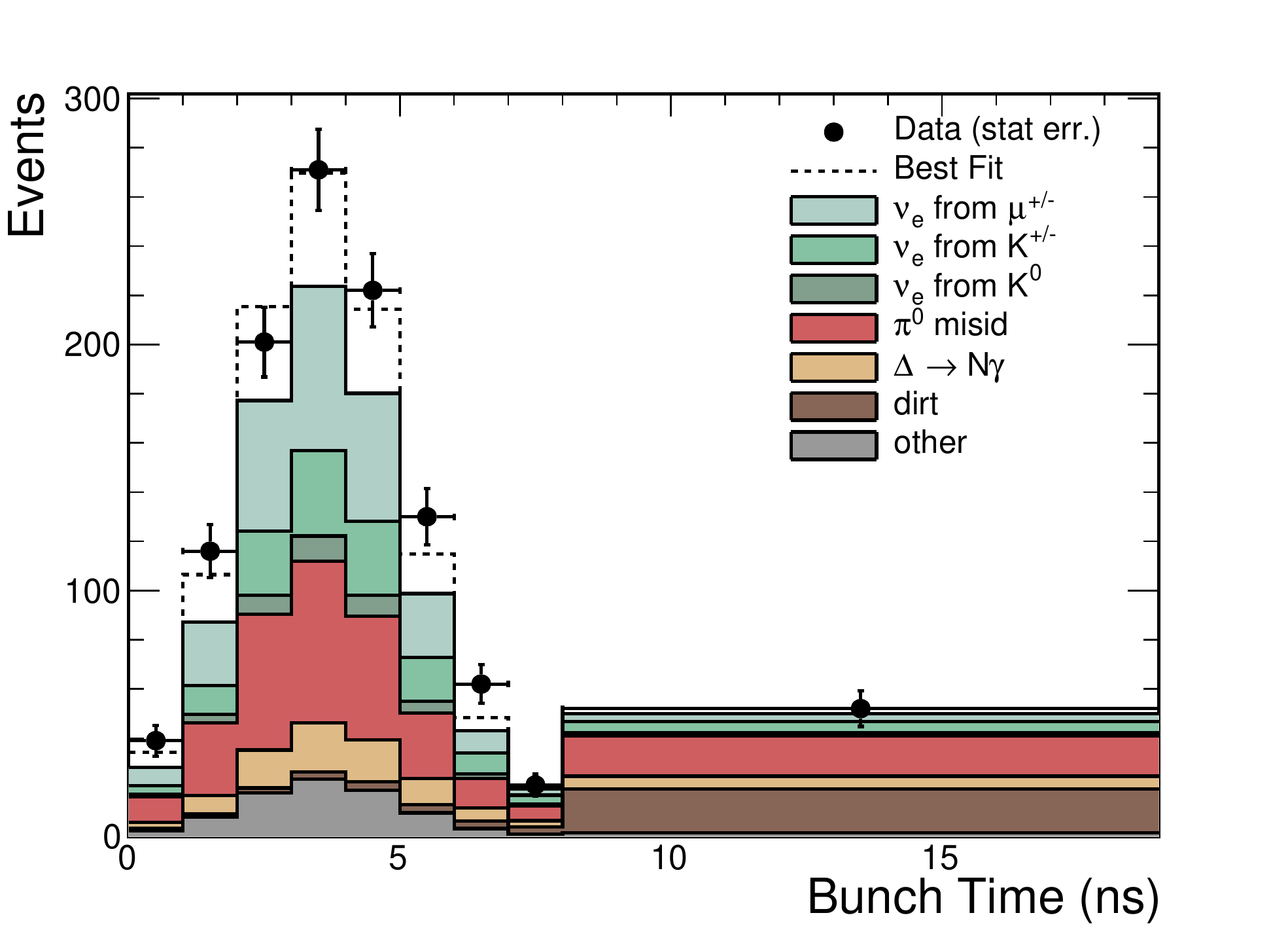}}
 \vspace{-0.3in}
\caption{The bunch timing for data events in neutrino mode compared to the expected background
in the $200<E_\nu^{QE}<1250$ MeV energy range. Almost all of the excess data
events occur, as expected from neutrino events in the detector, within the first 8 ns of the bunch timing.
This data sample uses events collected with the new fiber timing system and represents about 40\% of the entire neutrino mode sample.}
\label{Timing}
\vspace{0.1in}
\end{figure}

\bigskip

\begin{table}[t]
\vspace{-0.1in}
\caption{\label{Rfit} \em The result of log-likelihood shape-only fits to the radial
distribution in neutrino mode, 
assuming only statistical errors, where different processes are normalized to explain the observed event excess.
The two-neutrino hypothesis
fits the radial distribution best with a $\chi^2 = 8.4/9 ndf$, while the NC $\pi^0$ hypothesis has a worse fit
with a $\chi^2 = 17.2/9 ndf$. Also shown is the multiplicative factor that is required for each hypothesis 
to explain the observed event excess.
}
\small
\begin{ruledtabular}
\begin{tabular}{ccc}
Hypothesis&Multiplicative factor&$\chi^2 /9 ndf$ \\
\hline
NC $\Delta \rightarrow N \gamma$ Background & 3.18 & 10.0 \\
External Event Background & 5.98 & 44.9 \\
$\nu_e$ \& $\bar \nu_e$ from $K^0_L$ Decay Background & 7.85 & 14.8 \\
$\nu_e$ \& $\bar \nu_e$ from $K^{\pm}$ Decay Background & 2.95 & 16.3 \\
$\nu_e$ \& $\bar \nu_e$ from $\mu^{\pm}$ Decay Background & 1.88 & 16.1 \\
Other $\nu_e$ \& $\bar \nu_e$ Background & 3.21 & 12.5 \\
NC $\pi^0$ Background & 1.75 & 17.2 \\
Best Fit Oscillations & 1.24 & 8.4 \\
\hline
\end{tabular}
\vspace{-0.2in}
\end{ruledtabular}
\normalsize
\end{table}

\bigskip
\bigskip

Single-gamma backgrounds
from external neutrino interactions (``dirt" backgrounds) are estimated using topological and spatial cuts
to isolate the events whose vertices are near the edge of the detector and
point towards the detector center \cite{mb_lowe}.
The external event background estimate has been confirmed by measuring the absolute time of signal
events relative to the proton beam microstructure (52.81 MHz extraction frequency), which corresponds to buckets of beam approximately
every 18.9 ns. 
Fig. \ref{Timing} shows that the event excess peaks in the 8 ns window associated with beam bunch time, 
as expected from neutrino events in the detector, and is inconsistent with external neutrino events or 
beam-off events, which would be approximately flat in time. Also, the observed background level outside 
of the beam agrees well with the predicted background estimate. In addition, good agreement is obtained for the
event excess with $\cos \theta >0.9$.
The timing reconstruction performed here is similar to the reconstruction in
reference \cite{mb_dm}, but with a different time offset applied.

The $\Delta \rightarrow N + \gamma$ background is determined from the NC $\pi^0$ event sample \cite{mb_pi0}, which
has contributions from $\Delta$ production in $^{12}$C (52.2\%), $\Delta$ production in H$_2$ (15.1\%),
coherent scattering on $^{12}$C (12.5\%), coherent scattering on H$_2$ (3.1\%), higher-mass resonances
(12.9\%), and non-resonant background (4.2\%). The fraction of $\Delta$ decays to $\pi^0$
is 2/3 from the Clebsch-Gordon coefficients, and the probability of pion escape from the $^{12}$C nucleus
is estimated to be 62.5\%.
The $\Delta$ radiative branching fraction is 0.60\% for $^{12}$C and 0.68\% for H$_2$
after integration over all the invariant mass range, 
where the single gamma production branching ratio increases below the pion production threshold. 
With these values, the ratio of single gamma events to NC $\pi^0$ events, $R$, can
be estimated to be $$R=0.151 \times 0.0068 \times 1.5 + 0.522 \times 0.0060 \times 1.5 /0.625 = 0.0091.$$
Note that single gamma events are assumed to come entirely from $\Delta$ radiative decay. The total uncertainty
on this ratio is 14.0\% (15.6\%) in neutrino (antineutrino) mode. This estimate of $R=0.0091 \pm 0.0013$ 
agrees fairly well with theoretical calculations of the single gamma event rate \cite{hill_zhang}.

The intrinsic $\nu_e$ background comes almost entirely from muon and kaon 
decay-in-flight in the beam decay pipe. 
MiniBooNE $\nu_\mu$ CCQE event measurements \cite{mb_numuccqe}
constrain the size and energy dependence of the intrinsic $\nu_e$ background from muon decay, 
while the intrinsic $\nu_e$ background from kaon decay is constrained by fits to kaon production
data and SciBooNE measurements \cite{sciboone_kaon}. Furthermore, due to the higher energy of the intrinsic
$\nu_e$ background, this background is disfavored from the fit
to the radial distribution, as shown in Table \ref{Rfit}.

Finally, backgrounds from exotic $\pi^0$ decay in the neutrino production target are ruled out from
the MiniBooNE beam-dump run, where the incident proton beam was steered above the Be target and 
interacted in the steel beam dump at the downstream end of the decay pipe. No excess of events
was observed \cite{mb_dm}, which set limits on light dark matter and other exotic $\pi^0$ decays.

\section{Constraints on NC $\pi^0$ Background with Tighter Radius Selection}

\begin{figure}[tbp]
\vspace{-0.25in}
 \centerline{\includegraphics[angle=0,width=9.0cm]{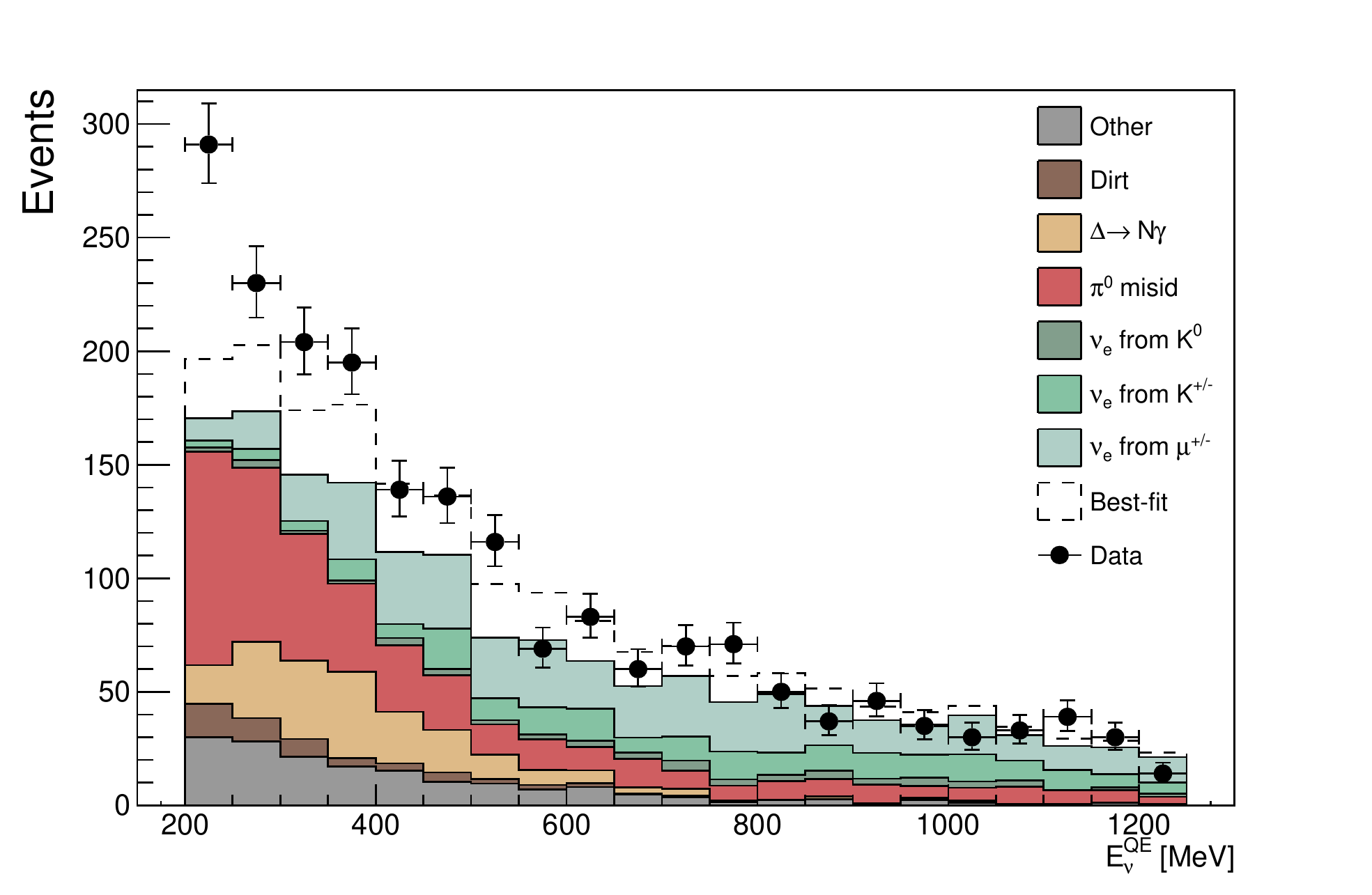}}
 \vspace{-0.3in}
\caption{The MiniBooNE neutrino mode
$E_\nu^{QE}$ distributions, corresponding to the total $18.75 \times 10^{20}$ POT data in neutrino mode
in the $200<E_\nu^{QE}<1250$ MeV energy range,
for ${\nu}_e$ CCQE data (points with statistical errors) and background
(histogram) with radius less than 4 m. The dashed histogram shows
the best fit to the neutrino-mode data assuming two-neutrino oscillations.}
\label{EnuQE_Rlt4}
\vspace{0.1in}
\end{figure}

\begin{figure}[tbp]
\vspace{-0.25in}
 \centerline{\includegraphics[angle=0,width=9.0cm]{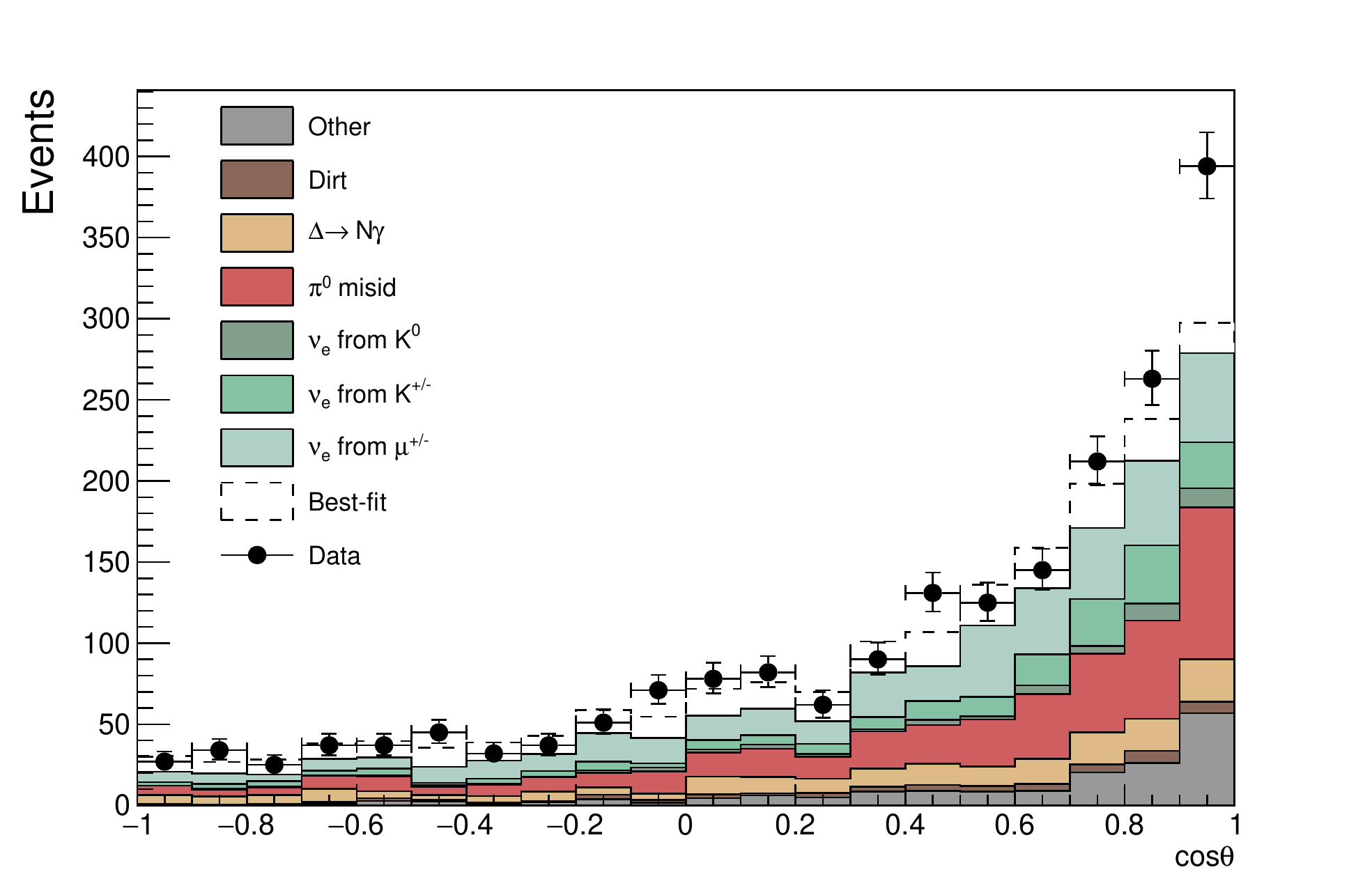}}
 \vspace{-0.3in}
\caption{The MiniBooNE neutrino mode
$\cos \theta$ distributions, corresponding to the total $18.75 \times 10^{20}$ POT data in neutrino mode
in the $200<E_\nu^{QE}<1250$ MeV energy range,
for ${\nu}_e$ CCQE data (points with statistical errors) and background
(histogram) with radius less than 4 m. The dashed histogram shows
the best fit to the neutrino-mode data assuming two-neutrino oscillations.}
\label{Uz_Rlt4}
\vspace{0.1in}
\end{figure}

Explanations for the event excess have included unsimulated photons entering the detector from external interactions and
the undersimulation of photons lost from $\pi^0$ production within the detector.   To test these explanations in a model-independent way,
we can use our higher event statistics to study the change in the excess as a function
of tighter fiducial volume cuts. The NC $\pi^0$ and external event backgrounds preferentially
populate higher radius compared to electron neutrino interactions. Therefore, reducing the fiducial radius is expected to
reduce the significance of the excess if it is due to these backgrounds and increase the significance of
the excess if its distribution is $\nu_e$-like. If we change the standard 5 m cut to 4 m, we find 
there are 1978 data events in neutrino mode, $1519.4 \pm 81.9$ background events, and an excess of 
$458.6 \pm 81.9$ events ($5.6 \sigma$). If we
use a 3 m cut, we find 864 data events, $673.9 \pm 41.2$ background events, and
an excess of $190.1 \pm 41.2$ events ($4.6 \sigma$), consistent with what is expected
if the signal is more $\nu_e$-like. The event statistics are shown in Table \ref{excesses},
while Figs. \ref{EnuQE_Rlt4} and \ref{Uz_Rlt4}
show the reconstructed neutrino energy and $\cos \theta$ distributions for electronlike events with radius less than 4 m for
both data events and background events. 

\section{Conclusion}

In summary, the MiniBooNE experiment 
observes a total excess 
of $638.0 \pm 52.1(stat.) \pm 132.8(syst.)$ electronlike events in the energy range $200<E_\nu^{QE}<1250$~MeV  
in both neutrino and antineutrino running modes.
The overall significance of the excess, $4.8 \sigma$, is limited by systematic uncertainties,
assumed to be Gaussian, as the statistical significance of
the excess is $12.2 \sigma$.
All of the major backgrounds are constrained by in situ 
event measurements.
Beam timing information shows that almost all of the excess is in time with neutrinos that interact 
in the detector. The radius distribution shows that the excess is distributed throughout the volume, 
while tighter cuts on the fiducal volume increase the significance of the excess. 
The data likelihood ratio disfavors models that
explain the event excess due to entering or exiting photons.
The MiniBooNE event excess will be further studied by the Fermilab 
short-baseline neutrino (SBN) program \cite{sbn} and by the JSNS$^2$ experiment at J-PARC \cite{jsns2}.

\begin{acknowledgments}
We acknowledge the support of Fermilab, the Department of Energy,
and the National Science Foundation, and
we acknowledge Los Alamos National Laboratory for LDRD funding. 
\vspace{-0.01in}
\end{acknowledgments}



\end{document}